\documentclass[aps,prd,secnumarabic,amssymb, amsmath,nobibnotes,nofootinbib,11pt]{revtex4} 
\usepackage{amsfonts,amsmath,hyperref,url, color}
\usepackage{bm, bbm}
\usepackage{graphicx}
\usepackage{mathtools}

\newcommand{\be}{\begin{equation}}
\newcommand{\bal}{\begin{align}}
\newcommand{\eal}{\end{align}}
\newcommand{\ee}{\end{equation}}
\newcommand{\bea}{\begin{eqnarray}}
\newcommand{\eea}{\end{eqnarray}}
\newcommand{\bit}{\begin{itemize}}
\newcommand{\eit}{\end{itemize}}

\newcommand{\ba}{\begin{aligned}}
\newcommand{\ea}{\end{aligned}}

\usepackage{color}

\begin{document}

\title{A new pairwise boost quantum number from celestial states}




\author{Francesco Alessio}
\email{francesco.alessio@su.se}
\affiliation{\it \small NORDITA, KTH Royal Institute of Technology and Stockholm University, \\
 Hannes Alfv{\'{e}}ns v{\"{a}}g 12, SE-11419 Stockholm, Sweden\\}
\affiliation{\it \small  Department of Physics and Astronomy, Uppsala University,\\ Box 516, SE-75120 Uppsala, Sweden}

\author{Michele Arzano}
\email{michele.arzano@na.infn.it}
\affiliation{Dipartimento di Fisica ``E. Pancini", Universit\`a di Napoli Federico II, I-80125 Napoli, Italy\\}
\affiliation{INFN, Sezione di Napoli,\\ Complesso Universitario di Monte S. Angelo,\\
Via Cintia Edificio 6, 80126 Napoli, Italy}

\begin{abstract}

Infrared effects in the scattering of particles in gravity and electrodynamics entail an exchange of relativistic angular momentum between pairs of particles and the gauge field. Due to this exchange particles can carry an asymptotically non-vanishing ``pairwise" boost-like angular momentum proportional to the product of their couplings to the field. At the quantum level this asymptotic angular momentum suggests the existence of a new quantum number carried by multi-particle states. We argue that such quantum number is related to a modification of the action of the generators of Lorentz transformations on multi-particle states. We derive such a modification using a group-theoretic argument based on the little group of the conformal primary basis for asymptotic states. The corresponding representation is an extension of the ordinary multi-particle Fock representation of the Poincar\'e group. The new multi-particle states belonging to such representation no longer factorize into tensor products of one-particle states. Viewed from a gravitational point of view, our results provide evidence for a universal breakdown of the description of multi-particle sates in terms of Fock space due to infrared back-reaction.

\end{abstract}

\maketitle

\section{Introduction}

More than fifty years ago, in a work \cite{Zwanziger:1972sx} that remained mostly unnoticed until recent (see e.g. \cite{Csaki:2020yei,Csaki:2022qtz,Csaki:2020inw}), Zwanziger showed that in the scattering of electrically and magnetically charged particles the electromagnetic field carries, asymptotically, a time independent angular momentum proportional to the following combination of the product of two different particles electric $e_i$ and magnetic $g_j$ charges: $\mu_{ij}\sim e_i g_j - g_i e_j$. 
In that work it was observed that the existence of such additional angular momentum requires a non-trivial action of the generators of Lorentz transformations 
on multiparticle states. This, in turn, leads to asymptotic scattering states no longer transforming like tensor products of free one-particle states i.e. according to the ordinary Fock representation. This fact was picked up recently in \cite{Csaki:2020yei} where the authors argued that the non-trivial transformation of asymptotic states can be described in terms of an additional quantum number associated to pairs of particles dubbed {\it pairwise helicity}.

Interestingly, as already noticed by Zwanziger in \cite{Zwanziger:1972sx}, there are potential additional Coulombic contributions to the asymptotic angular momentum of the electromagnetic field proportional to the product of electric and magnetic charges of pairs of particles. 
Zwanziger dismissed these contributions arguing that they should vanish. Recent works \cite{Gralla:2021eoi,Bhardwaj:2022hip}, however, suggest that  these contributions should be present even in the case of electromagnetism without magnetic charges. 
In \cite{Gralla:2021eoi} the authors calculated the balance of four-momentum and relativistic angular momentum for the scattering of charged particles at leading order in the interaction. Their result shows that there is a shift of the net boost-like angular momentum or {\it mass moment} of the particles which is balanced by an equal and opposite change of the boost-like angular momentum carried by the field. They dubbed such effect {\it electromagnetic scoot}. The same contributions to the boost-like angular momentum of the particles and electromagnetic field were later re-derived in \cite{Bhardwaj:2022hip} in a covariant fashion along the lines of the orginal calculation of Zwanziger in \cite{Zwanziger:1972sx}.

The authors of \cite{Gralla:2021eoi} found also\footnote{Actually the authors discovered the gravitational scoot before the electromagnetic one.} that a similar effect arises in the gravitational scattering of massive particles \cite{Gralla:2021qaf}. In this case the gravitational scoot is proportional to the Newton's constant and the product of the two particles masses. These observations suggest that the original argument of Zwanziger in \cite{Zwanziger:1972sx} (and its modern reinterpretation given in \cite{Csaki:2020yei}) concerning the inadequacy of a Fock representation for free particle states can be extended to {\it ordinary} electromagnetism (i.e. without resorting to the existence of magnetic charges) and, most importantly, to gravity. Given the universal coupling of the latter to all forms of matter and enegy this strongly hints at an inadequacy of the Fock representation for free particles particles when gravity is switched-on. The root of such inadequacy lies in the presence of an additional quantum number, a {\it pairwise boost-like angular momentum}, associated to an infrared back-reaction effect sourced by soft gravitons. Indeed, as we discuss, the presence of a non-vanishing scoot can be related to logarithmic terms in the subleading soft graviton and electromagnetic theorems, as extesively discussed in \cite{Laddha:2018myi,Sahoo:2018lxl,Saha:2019tub,Sahoo:2020ryf,Krishna:2023fxg}.

In this work we provide group-theoretic evidence for the existence of this new pairwise quantum number. Our argument relies on the conformal primary basis for asymptotic scattering states introduced in \cite{Pasterski:2016qvg,Pasterski:2017kqt} which play a pivotal role in celelestial holography \cite{Pasterski:2021rjz,Pasterski:2021raf,Arkani-Hamed:2020gyp,Raclariu:2021zjz,Iacobacci:2024nhw,Iacobacci:2020por,Sleight:2023ojm,Ball:2019atb,Pate:2019lpp,Fotopoulos:2019tpe,Law:2020tsg,Fotopoulos:2020bqj}. Such states are labelled by coordinates on the celestial sphere and a conformal dimension which is the eigenvalue of the generator of boosts in a given direction. When considering a two-particle system we point out that, besides rotations along the axis of motion of the two-particles already discussed by Zwanziger in \cite{Zwanziger:1972sx}, the little group can be consistently enhanced in order to include an additional Lorentz transformation preserving the two-particle configuration in the conformal primary representation: boosts along the direction of motion of the two particles. It is precisely the eigenvalue associated with such transoformation which will account for the pairwise boost-like angular momentum discovered in \cite{Gralla:2021qaf,Gralla:2021eoi}.

The paper is organized as follows: in the next section we briefly recall recent results on the existence of an symptotic boost-like angular momentum in particle scattering in gravity and electromagnetism. Section \ref{Sec3} is devoted to an introduction and review of massive conformal primary states. In particular, in \ref{Sec3.1} we introduce the basis in terms of the standard on-shell momentum basis and fix our conventions. In \ref{Lor} we discuss the action of Lorentz transformations on such basis and in \ref{Sec3.3 } we derive the form of their generators both in the on-shell momentum and conformal primary basis. In Section \ref{Sec4} we discuss an enhancement of the standard Wigner little group for massive multiparticle states to include boosts in the direction of motion of then particles. We first discuss how such the pairwise boost quantum number is related to the little group of the conformal primary states and then extend the derivation to multiparticles states described in the ordinary on-shell four-momentum basis. This allows us to relate the new pairwise boost quantum number to the scoot effect discussed in Section \ref{sec1}. We conclude with a summary of the results presented and a brief discussion of future perspectives.
 
Throughout the paper we use mostly plus signature.

\section{IR scoot in gravity and electrodynamics}
\label{sec1}
Motivated by the study of gravitational self-force effects in the post-Minkowskian (PM) regime, the authors of \cite{Gralla:2021qaf} derived a new formula for the shift of the total mass moment (or \textit{boost charge}) $N^i$ of a system of two point particles involved in a gravitational scattering process. As remarked in the introduction, such quantity is defined as the boost-like component of the total mechanical angular momentum of the system, 
\begin{align}
\label{F1}
M^{\mu\nu}=\sum_{a=1,2}M^{\mu\nu}_{a}=\sum_{a=1,2}p_a^{\mu}x^{\nu}_{a}-p_a^{\nu}x^{\mu}_{a},\qquad N^i\equiv M^{0i}=\sum_{a=1,2}E_ax_a^i-t\hspace{0.05cm}p_a^{i},
\end{align}
where $p_a^{\mu}=(E_a,p_a^i)$ and $x_a^{\mu}=(t,x^i)$ are the momenta and the positions of the particles involved in the scattering process.
In particular, in the rest frame of particle 2 the position four-vectors are $\bar{x}^{\mu}_1=(t,b,0,vt)$ and $ \bar{x}^{\mu}_2=(t,0,0,0)$, whereas their initial momenta are,
\begin{align}
\label{F2}
\bar{p}_1^{\mu}=m_1\gamma(1,0,0,v),\qquad \bar{p}_2^{\mu}=(m_2,0,0,0),\qquad\gamma=-\frac{\bar{p}_1\cdot \bar{p}_2}{m_1 m_2}=\frac{1}{\sqrt{1-v^2}},
\end{align}
where $v$ and $b$ are their relative velocity and position, namely the impact parameter, and where $\gamma$ is the usual Lorentz factor. It can be immediately checked that in this frame $\bar{N}^i_{\mathrm{mec}}$ does not vanish. However, in the center of energy-momentum (CEM) frame,
defined by the Poincar\'e transformation,
\begin{align}
\label{CEM1}
x_i^{\mu}=\Lambda^{\mu}{}_{\nu}\bar{x}_i^{\nu}+a^{\mu}\qquad p_i^{\mu}=\Lambda^{\mu}{}_{\nu}\bar{p}_i^{\nu},
\end{align}
where,
\begin{align}
\label{CEM2}
\Lambda^{\mu}{}_{\nu}=\left(\begin{matrix}\frac{m_2+\gamma m_1}{E_0}& 0 & 0 & -\frac{\gamma m_1 v}{E_0}\\ 0 & 1 & 0 & 0 \\0 & 0 & 1 & 0\\-\frac{\gamma m_1 v}{E_0}& 0 & 0 & \frac{m_2+\gamma m_1}{E_0}\end{matrix}\right),\qquad a^{\mu}=-b\frac{m_1(\gamma m_2+m_1)}{E_0^2}\delta^{\mu}_{1},
\end{align}
and,
\begin{align}
E_0=\sqrt{-(p_1+p_2)^2}=\sqrt{m_1^2+m_2^2+2m_1m_2\gamma},
\end{align}
is the total energy, we have that the total spatial momentum is zero, $P^i_{\mathrm{mech}}=p_1^{i}+p_2^{i}=0$ and also that $N^i_{\mathrm{mech}}=p_1^{i}+p_2^{i}=0$. 
Explicitly for the momenta we have,
\begin{align}
\label{CEM2}
p_1^{\mu}=\frac{m_1\gamma}{E_0}(m_2-m_1\gamma(v^2-1)
,0,0,m_2v),\qquad p_2^{\mu}=\frac{m_2}{E_0}(m_2+\gamma m_1,0,0,-m_1v\gamma).
\end{align}
In this frame, it was found that the aforementioned shift is explicitly given by the following formula,
\begin{align}
\label{F3}
\Delta N^z= \frac{2Gm_1m_2\gamma(3-2\gamma^2)}{\gamma^2-1}\log\left(\frac{m_2+\gamma m_1}{m_1+\gamma m_2}\right)\,.
\end{align}
Note that the numerator in this formula is the same of the classical subleading soft factor discussed e.g. in the last line of (3.6) of \cite{Sahoo:2018lxl}, where $G$ is the gravitational constant. This effect has been referred to as \textit{gravitational scoot}. Being linear in $G$, from the PM expansion point of view \cite{Kosower:2018adc,Cristofoli:2021vyo,DiVecchia:2023frv,Kosower:2022yvp,Buonanno:2022pgc,Travaglini:2022uwo}, the above formula can be interpreted as a 1PM contribution to the the radiated angular momentum during the scattering process. The  2PM (one-loop) order \cite{Manohar:2022dea,DiVecchia:2022owy,Alessio:2022kwv} has been shown to be entirely due to the radiation of soft (\textit{i.e.} zero frequency) gravitons. Note that while the PM expansion assumes a small dimensionless quantity $Gm/b$, where $m$ is one of the two particles masses, equation \eqref{F3} is actually independent of the impact parameter $b$ of the process, which drops in the final result. Therefore, the gravitational scoot is universal in the sense that it does not depend on the details of the scattering process and it is a consequence of the long-range nature of the gravitational interaction that influences the asymptotic behaviour of the particle trajectories, which is not the same as that of free particles. Indeed, as already pointed out in \cite{Zwanziger:1973if} in the context of electromagnetism, two free particle trajectories due to similar infrared effects acquire a correction which depends logarithmically on their proper time. On these lines of thought, the authors of \cite{DiVecchia:2022owy} conjectured that this contribution could be captured by the subleading soft graviton (or photon) theorem \cite{Cachazo:2014fwa,Bern:2014vva,Bern:2014oka}. The latter indeed plays an important role in celestial holography and its connection to asymptotic symmetries \cite{Compere:2018ylh,Donnay:2020guq,Donnay:2022hkf,Agrawal:2023zea,Strominger:2013jfa,He:2014laa,Strominger:2017zoo,Cachazo:2014fwa,Alessio:2019cae} as well as in the context of amplitudes for classical gravitational scattering \cite{DiVecchia:2022owy,DiVecchia:2021ndb,Alessio:2022kwv,Alessio:2024wmz,Lippstreu:2023vvg,McLoughlin:2022ljp}.
\\

In the easier case of electromagnetism, which already captures the relevant structures of the problem, a similar formula was found in  \cite{Gralla:2021eoi} for the shift in the total boost charge of a two-particle system:
\begin{align}
\label{F4}
\Delta N^z= \frac{2e_1e_2}{\gamma^2-1}\log\left(\frac{m_2+\gamma m_1}{m_1+\gamma m_2}\right),
\end{align}
where now $e_1$ and $e_2$ are the electric charges of the particles. In this context, it has been demonstrated \cite{Gralla:2021eoi} that the mechanical mass moment in \eqref{F4} is exactly compensated by an equal and opposite mass moment shift of the electromagnetic field $\Delta N^z_{\mathrm{em}}$:
\begin{align}
\label{F5}
\Delta N^z+\Delta N^z_{\mathrm{em}}=0,\qquad N^i_{\mathrm{em}}=\int d^3x\hspace{0.05cm}\mathcal{E} x^i-t\int d^3x \mathcal{P}^i, 
\end{align}
where $\mathcal{E}$ and $\mathcal{P}^i$ are the field energy and momentum densities. This shows that there is an exchange between mechanical and field degrees of freedom. One concludes that even if it is possible to set $N^z=0$ at early times by the coordinate transformation in \eqref{CEM1}, the total boost charge at late times, after the scattering process, will be non-vanishing.

The shift in the boost charge can be easily derived from the corrected equations of the worldlines describing the asymptotic motion for large values of the proper time of two classical relativistic particles under the mutual influence of their Coulomb fields \cite{Zwanziger:1972sx,Sahoo:2018lxl} 
\begin{align}
\label{log1}
&x^{\mu}_a(\tau_a)=\frac{p^{\mu}_a}{m_a}\tau_a+\eta_a\log|\tau_a|\sum_{b\neq a}e_ae_b\frac{\gamma u^{\mu}_a-u_b^{\mu}}{m_a(\gamma^2-1)^{\frac{3}{2}}},\\
\label{log1.5}
&p^{\mu}_a(\tau_a)=p^{\mu}_a+\eta_a\frac{1}{\tau_a}\sum_{b\neq a}e_ae_b\frac{\gamma u^{\mu}_a-u_b^{\mu}}{(\gamma^2-1)^{\frac{3}{2}}},
\end{align}
where $\eta_a=1$  $(-1)$  for outgoing (incoming) particles and $\tau_a$ is the proper time of the $a$-th particle. 
The $\log$ dependence in the position four-vectors is an unavoidable consequence of the long-range nature of the Coulomb force, for an inverse-square force integrates up to a logarithmically divergent position. Similarly, such logarithms appear already in the tree-level elastic eikonal \cite{DiVecchia:2023frv,Alessio:2024wmz} in four-dimensions.

 We denote by $\Delta M^{\mu\nu}$ the net difference between the asymptotic angular momentum at late and early times and we assume that the late and early time cutoffs for the particles are $\tau_a^{+}$ and $\tau_a^{-}$, respectively. The asymptotic angular momentum of the $a$-th particle is
 \begin{align}
 \label{angmom}
M^{\mu\nu}_a(\tau^{\eta_a}_a)=-2\eta_a\sum_{b\neq a}\frac{e_a e_b}{(\gamma^2-1)^{\frac{3}{2}}}u^{[\mu}_au^{\nu]}_b\log|\tau_a^{\eta_a}|
 \end{align}
 and hence
\begin{align}
\label{log3}
\Delta M^{\mu\nu}&=\sum_{a}M^{\mu\nu}(\tau_a^+)-M^{\mu\nu}(\tau_a^-)=-2\sum_{a}\sum_{b\neq a}\frac{e_ae_b u_a^{[\mu}u_b^{\nu]}}{(\gamma^2-1)^{\frac{3}{2}}}\log|\tau_a^+\tau_a^-|.
\end{align}
If we further assume that $\tau_a^+=\tau_a^-\equiv \tau_a$,  
we have
\begin{align}
\label{log3.5}
\Delta M^{\mu\nu}=-4\sum_{a}\sum_{b\neq a}\frac{e_ae_b u_a^{[\mu}u_b^{\nu]}}{(\gamma^2-1)^{\frac{3}{2}}}\log|\tau_a|.
\end{align}
The above equation could also be inferred from the analysis in \cite{Bhardwaj:2022hip}. When specifying to the case of a two-particle scattering, plugging in the previous formula the explicit parametrization of the momenta \eqref{CEM2} in the CEM frame we get
\begin{align}
\label{match}
\Delta N^z= \frac{2e_1e_2}{\gamma^2-1}\log\bigg|\frac{\tau_1}{\tau_2}\bigg|,
\end{align}
matching equation \eqref{F4} upon identification of the arguments of the logarithms. The actual value of the latter is not important in the present analysis and it depends on the initial conditions of the scattering problem which we did not include in the trajectories displayed in \eqref{log1} and \eqref{log1.5}. 

Similarly, the asymptotic trajectories of particles subjected to their reciprocal gravitational interaction are given by \cite{Sahoo:2018lxl}
\begin{align}
\label{gr1}
&x_a^{\mu}(\tau_a)=\frac{p_a^{\mu}}{m_a}\tau_a+\eta_a\log|\tau_a|\sum_{b\neq a}Gm_b\frac{u_a^{\mu}+ u_b^{\mu}\gamma(2\gamma^2-3)}{(\gamma^2-1)^{\frac{3}{2}}},\\&
p_a^{\mu}(\tau_a)=p_a^{\mu}+\eta_a\frac{1}{\tau_a}\sum_{b\neq a}G m_b\frac{u_a^{\mu}+u_b^{\mu}\gamma (2\gamma^2-3)}{(\gamma^2-1)^{\frac{3}{2}}}.
\end{align}
Proceeding as before, we get
for the asymptotic angular momentum of outgoing/incoming particles
\begin{align}
M^{\mu\nu}_a(\tau_a)=2\eta_aG\sum_{b\neq a}\frac{m_am_bu_a^{[\mu}u_b^{\nu]}\gamma(2\gamma^2-3)}{(\gamma^2-1)^{\frac{3}{2}}}\log|\tau_a|,
\end{align}
and
\begin{align}
 \Delta M^{\mu\nu}=4G\sum_{a}\sum_{b\neq a}\frac{m_am_bu_a^{[\mu}u_b^{\nu]}\gamma(2\gamma^2-3)}{(\gamma^2-1)^{\frac{3}{2}}}\log|\tau_a|.
\end{align}
for the shift.
In the CEM frame and for two particles we recover \eqref{F3}, as we should:
\begin{align}
\label{grsc}
\Delta N^z=\frac{2Gm_1m_2\gamma(3-2\gamma^2)}{\gamma^2-1}\log\left|\frac{\tau_1}{\tau_2}\right|.
\end{align}
In the rest of this work we argue that the existence of this asymptotic boost charge is reflecte into a modification of the generator of boosts on multi-particle states described in the conformal primary representation which we introduce in the next section.

\section{Conformal primary basis}
\label{Sec3}
The Wigner little group of massive particles, \textit{i.e.} the subset of the Lorentz group that leaves a reference four-momentum invariant is that of three-dimensional spatial rotations, $\mathrm{SO}(3)$, whose associated quantum number $\sigma$ is the spin labelling massive irreducible representations of the Lorentz group. Correspondingly, single-particle states are represented by kets $|p,\sigma\rangle$, where $p$ is the on-shell four-momentum of the particle.
Multiparticle states are customarily constructed as ``bare" tensor products of single-particle states such as $|p_1,p_2,\sigma_1,\sigma_2\rangle=|p_1,\sigma_1\rangle\otimes|p_2.\sigma_2\rangle$. However, as it is clear from the discussion in the previous section, when considering the interaction of massive particles mediated by electromagnetic or gravitational interaction, the presence of a asymptotically non-vanishing boost-like angular momentum coupling pairs of particles makes this simple construction inadequate. In order to address such 
shortcoming, as a first step, we look for a basis of asymptotic multiparticle states characterized by a boost quantum number 
i.e. which diagonalizes the action of boosts. This is achieved by the conformal primary basis \cite{Pasterski:2016qvg,Pasterski:2017kqt} which we introduce here in the case of a scalar massive particle. A generalization for particles with spin can be found e.g. in \cite{Pasterski:2020pdk}.

\subsection{The basis}
\label{Sec3.1}
We start by considering the Minkowski metric in momentum space
\begin{align}
    \label{F14}
ds^2=-dE^2+d\vec{p}^2=-dE^2+dp^2+p^2(d\theta^2+\sin\theta^2d\phi^2).
\end{align}
The induced metric on the mass-shell hyperboloid $H_3$ defined by $E^2=p^2+m^2$ for a fixed value of the mass $m$ is
\begin{align}
    \label{F15}
    ds^2=m^2ds^2_{H_3}, \qquad ds^2_{H_3}=\frac{dy^2+dz d\bar z}{y^2},
\end{align}
where $ds^2_{H_3}$ is the metric on the unit mass hyperboloid and the Poincar\'e coordintes $(y,z,\bar{z})$ are defined as
\begin{align}
\label{F16}
y=\frac{m}{\sqrt{m^2+p^2}+p\cos\theta},\qquad z=\frac{p \sin\theta}{\sqrt{p^2+m^2}+p\cos\theta}e^{i\phi},\qquad \bar{z}=z^*\,.
\end{align}
Notice that the conformal boundary is located at $y=0$ and it is conformally a 2-sphere $S^2$.

The on-shell four-momentum of a massive particle can be thus parametrized \cite{Raclariu:2021zjz, Pasterski:2021rjz} in terms of such coordinates as\footnote{We will use the shorthand notation $f(y,z,\bar z)\equiv f(y,z)$.}
\begin{equation}\label{momzy}
    p^{\mu}(y,z) =\eta\frac{m}{2y} \left(1+y^2+z\bar{z} ,z+\bar{z},i (\bar{z}-z),1-y^2-z\bar{z}\right) \equiv \eta m\, \hat{p}^{\mu}(y,z),
\end{equation}
with $\hat{p}^2=-1$ and $\eta=\pm 1$ for outgoing/incoming particles as in the previous section. One-particle states of scalar particles in the $(y,z,\bar z)$ momentum basis are taken to be $|p\,\rangle \equiv |y,z\rangle$ and they are associated to plane wave solutions $\langle x|y,z\rangle$ of the Klein-Gordon equation. Completeness of the $|y,z\rangle$ basis implies that
\begin{align}
|y,z\rangle=\int d\mu(p')\langle y',z'|y,z\rangle|y',z'\rangle=\int_0^{\infty} \frac{d y'}{y'^3} \int dz' d\bar{z}'{}^{\eta}\langle y',z'|y,z\rangle|y',z'\rangle,
\end{align}
where $d\mu(p)$ is the Lorentz invariant integration measure on $H_3$, 
$d\mu(p) =\frac{d^3 \hat{p}}{\hat{p}^0}$,
from which we read the orthonormality condition
\begin{align}
\label{ort0}
\langle y',z'|y,z\rangle=y^3\delta(y-y')\delta^{(2)}(z-z').
\end{align}
The  {\it conformal primary basis} $|\Delta,w,\bar w\rangle \equiv|\Delta,w\rangle$ is defined as \cite{Lam:2017ofc}
\begin{align}
\label{cfb}
|\Delta,w\rangle=\int d\mu(p)\langle y,z|w,\Delta\rangle |y,z\rangle\equiv \int_0^{\infty} \frac{d y}{y^3} \int dz d\bar{z}G_{\Delta}(\hat{p},w)|y,z\rangle,
\end{align} 
where the coefficients of the integral map are given by the scalar bulk-to-boundary propagator on $H_3$ \cite{Witten:1998qj,Pasterski:2017kqt}
\begin{equation}
    G_{\Delta}(\hat{p},w) \equiv \langle y,z|\Delta,w\rangle = \frac{1}{\sqrt{4\pi^3}}\frac{\Gamma(1+i\lambda)}{\Gamma(i\lambda)}\bigg(\frac{y}{y^2+|w-z|^2}\bigg)^{\Delta},\qquad \Delta= 1+i\lambda.
\end{equation}
$\Delta$ is the {\it conformal dimension}. The states are orthonormal
\begin{align}
\label{ort}
\langle \Delta',w'|\Delta,w\rangle=\int_0^{\infty} \frac{d y}{y^3} \int dz d\bar{z}G^{*}_{\Delta'}(\hat{p},w')G_{\Delta}(\hat{p},w)=\delta(\lambda-\lambda')\delta^{(2)}(w-w'),
\end{align}
where we used \eqref{ort0} and 
the value of the integral\footnote{We thank the authors of \cite{Sleight:2023ojm} for bringing \cite{Costa:2014kfa} to our attention.}\cite{Costa:2014kfa}
\begin{equation}
\label{int}
\int \frac{dy}{y^{d+1}}\int d^d\vec{z}\frac{y^{\Delta+\Delta'+\epsilon}}{[y^2+(\vec{z}-\vec{w}')^2]^{\Delta'}[y^2+(\vec{z}-\vec{w})^2]^{\Delta}
}=\frac{2\pi^{d+1}|\Gamma(i\lambda)|^2}{|\Gamma(\frac{d}{2}+i\lambda)|^2} \delta(\lambda+\lambda') \delta^{(d)}(\vec{w}-\vec{w}').
\end{equation}
where $(\Delta,\Delta')=(\frac{d}{2}+i\lambda,\frac{d}{2}+i\lambda')$.
The integral transform in \eqref{cfb} can also be inverted and states $|y,z\rangle$ in the on-shell momentum basis can be obtained from the conformal primary ones $|\Delta,w\rangle$ as
\begin{align}
\label{momb}
|y,z\rangle=\int
d\mu(\lambda,w)\langle \Delta,w|y,z\rangle|\Delta,w\rangle\equiv\int_0^{\infty} d\lambda\int dwd \bar w G^{*}_{\Delta}(\hat{p};w)|\Delta,w\rangle,
\end{align}
and the orthonormality condition in \eqref{ort0} implies that
\begin{align}
\label{ort3}
\int_0^{\infty}d\lambda\int dwd\bar w G^*_{\Delta}(\hat{p},w)G_{\Delta}(\hat{p}',w)=y^3\delta(y-y')\delta^{(2)}(w-w').
\end{align}
In the momentum basis the eigenfunctions are simply given by standard plane waves,
\begin{align}
 f(x;y,z)\equiv \langle x|y,z\rangle=\frac{1}{(2\pi)^{\frac{3}{2}}}e^{i\eta  m\hat{p}(y,z)\cdot x}.
\end{align}
In the conformal primary basis they can be explicitly constructed by inserting the identity as
\begin{align}
\label{cppw}
\nonumber\phi_{\Delta}(x;w)&\equiv \langle x|\Delta,w\rangle=\int d^3\mu(p)\langle y,z|w,\Delta\rangle\langle x|y,z\rangle\\&=\frac{1}{(2\pi)^{\frac{3}{2}}}\int_0^{\infty}\frac{dy}{y^3
}\int dz d\bar z G_{\Delta}(\hat{p},w) e^{ i\eta m\hat{p}(y,z)\cdot x}.
\end{align}
For the sake of completeness the explicit form of the conformal primary wavefunctions in \eqref{cppw} is given by \cite{Pasterski:2017kqt,Pasterski:2020pdk,Raclariu:2021zjz,Iacobacci:2020por}
\begin{align}
\label{wavefct}
\phi_{\Delta}(x;w)=\frac{\Gamma(1+i\lambda)}{\Gamma(i\lambda)}\frac{1}{\sqrt{2}\pi^2}
\frac{1}{im}\frac{(\sqrt{-x^2})^{\Delta-1}}{(-q(w)\cdot x-i\eta \epsilon)^{\Delta}}K_{\Delta-1}(im\sqrt{-x^2}),
\end{align}
where $q^{\mu}(w)$ is the null vector  
\begin{align}
\label{nullmomx}
q^{\mu}(w)=(1+|w|^2,w+\bar w,i(\bar w-w),1-|w|^2),\qquad q^2=0,
\end{align}
and where $K_{\nu}(x)$ are modified Bessel function of the second kind. The wave functions in \eqref{wavefct} comprise a complete set of orthonormal solutions to the massive Klein-Gordon equation provided that $\lambda\in\mathbb R^{+}$ so that $\Delta$ belongs to principal continuous series of the irreducible representations of the Lorentz group \cite{Pasterski:2017kqt}. Finally, we also notice that, in terms of $q^{\mu}(w)$, the bulk-to-boundary propagator admits a simple representation,
\begin{align}
G_{\Delta}(\hat{p},w) = \frac{1}{\sqrt{4\pi^3}}\frac{\Gamma(1+i\lambda)}{\Gamma(i\lambda)}\frac{1}{[-\hat{p}(y,z)\cdot q(w)]^{\Delta}}.
\end{align}
\subsection{Action of Lorentz transformations}
\label{Lor}
We are now interested in the transformation properties of the new basis under the action of SL$(2,\mathbb{C})/\mathbb{Z}_2$. As shown in appendix \ref{appA}, on-shell momenta \eqref{momzy} transform under the action of SL$(2,\mathbb{C})/\mathbb{Z}^2$ through the coordinate map
\begin{equation}
\label{coordtrans}
y\rightarrow y' = \frac{y}{|cz+d|^2+|c|^2y^2}\equiv\Lambda y,\qquad z\rightarrow z' = \frac{(az+b)(\bar{c}\bar{z}+\bar{d})+a\bar{c}y^2}{|cz+d|^2+|c|^2y^2}\equiv \Lambda z, \qquad ,
\end{equation}
where $ad-bc=\bar{a}\bar{d}-\bar{b}\bar{c}=1$. The $|\Delta, w\rangle$ basis is defined through the integral transform \eqref{cfb} involving the bulk-to-boundary propagator. In the massless case, due to the nature of null hypersurfaces,  $(w,\bar w)$ and $(z,\bar z)$ are the same angular coordinates on both bulk and asymptotic spheres, defined through the standard stereographic map. However, in the massive case $(w,\bar w)$ have \textit{not} to be interpreted as the same as $(z,\bar z)$. Rather, the full set of variables $(\Delta, w,\bar w)$ has to be regarded as variables ``dual" to $(y, z,\bar z)$. As discussed in \cite{Pasterski:2016qvg} , trajectories of free massive particles $x^{\mu}(s)=\hat{p}^{\mu}s$ at early and late times can be written as $x^{\mu}(s)/\sqrt{-x^2(s)}\rightarrow \hat{p}^{\mu}$. Worldlines of massive particles thus asymptote to a fixed position on the hyperbolic slices of Minkowski space-time determined by their four-momenta. Hence $(w,\bar w)$ can be interpreted as a boundary coordinate of the early- and late-time asymptotic $H_3$ slice and can be considered as coordinates on past and future timelike infinity $i^{\mp}$, which are indeed conformal spheres. Hence, their transformation follows from evaluating \eqref{coordtrans} at $y=0$ and it reads 
\begin{align}
\label{mobius}
    w\rightarrow w'=\frac{aw+b}{cw+d}\equiv\Lambda w,\qquad ad-bc=1,
\end{align}
i.e. they undergo standard M\"obius transformations. By explicit computation, we can show the bulk-to-boundary propagator transforms with conformal weight $\Delta$
\begin{align}
    \label{F20}
    G_{\Delta}(\Lambda\hat{p},\Lambda w)=|cw+d|^{2\Delta}G_{\Delta}(\hat{p},w).
\end{align}
We can easily derive the transformation law $U(\Lambda)|\Delta,w\rangle$. We \textit{assume} that
\begin{align}
    \label{tr7}
    U(\Lambda)|\Delta,w\rangle \equiv f(\Lambda;w)|\Delta,\Lambda w\rangle,
\end{align}
with $f(\Lambda;w)$ to be determined. Then we have,
\begin{align}
\nonumber
    \label{tr8}
    \langle y,z|\Delta,w\rangle&=G_{\Delta}(\hat{p},w)\stackrel{\eqref{F20}}{=}G_{\Delta}(\Lambda\hat{p},\Lambda w)|cw+d|^{-2\Delta}=\langle \Lambda y,\Lambda z|\Delta,\Lambda w\rangle|cw+d|^{-2\Delta}\\&\stackrel{\eqref{tr7}}{=}\langle y,z|\underbrace{U^{\dagger}(\Lambda)U(\Lambda)}_{\mathbf{1} }|\Delta,w\rangle\frac{|cw+d|^{-2\Delta}}{f(\Lambda;w)}=G_{\Delta}(\hat{p},w)\frac{|cw+d|^{-2\Delta}}{f(\Lambda;w)}.
\end{align}
Therefore, comparing the last to the first line we must have $f(\Lambda;w)=|cw+d|^{-2\Delta}$ and thus
\begin{align}
    \label{F21}
U(\Lambda)|\Delta,w\rangle=|cw+d|^{-2\Delta}|\Delta,\Lambda w\rangle\,,
\end{align}
as also discussed in \cite{Banerjee:2018gce,Lippstreu:2021avq} in the massless case. Our equation \eqref{F21} shows that the same considerations of these works can be extended to the massive case. In particular, if we take $|\Delta,0\rangle$ as the reference state associated to a particle moving along the $x^3$ axis, its little group, together with rotations about the $x^3$ axis, is enhanced to include  boosts along the $x^3$ direction. Indeed, it is easy to prove that rotations and boosts along $x^3$ map $w=0$ into itself through \eqref{mobius}.

The transformation of the conformal primary wavefunctions \eqref{wavefct} follows easily from \eqref{F21}. We have
\begin{align}
    \label{gen1}
    \phi_{\Delta}(\Lambda\cdot x;\Lambda w)=\int\frac{d^3\hat{p}}{\hat{p}^0}G_{\Delta}(\hat{p},\Lambda w)e^{i\eta m\hat{p}\cdot\Lambda\cdot x}=\int\frac{d^3\hat{p}'}{\hat{p}'^0}G_{\Delta}(\hat{p}',w')e^{ i\eta m\hat{p}'\cdot x'}.
\end{align}
where we defined $x'\equiv\Lambda\cdot x$ and we changed the integration variable to $p'=\Lambda \cdot p$ and used Lorentz invariance of the measure $d\mu(p)$. Using now \eqref{F20} we obtain,
\begin{align}
    \label{gen1.1}
     \phi_{\Delta}(\Lambda\cdot x;\Lambda w)&=|cw+d|^{2\Delta}\int\frac{d^3\hat{p}'}{\hat{p}'^0}G_{\Delta}(\hat{p},w)e^{i\eta m\hat{p}'\cdot x'}=|cw+d|^{2\Delta}\phi_{\Delta}(x,w),
\end{align}
where in the last step we used again Lorentz invariance of the measure and that $p'\cdot x'=p\cdot x$. In the case of wavefunctions with spin, \textit{i.e.} for fields $\phi_{h,\bar h}(x,w)=\langle x|h,\bar h,w\rangle$ with $(h,\bar h)\equiv\frac{1}{2}(\Delta+\sigma,\Delta-\sigma)$ with  $h\neq \bar h$ and $\sigma$ is the spin, the transformations for the states and the wavefunctions are \cite{Pasterski:2020pdk},
\begin{align}
\label{spin1}
&U(\Lambda)|h,\bar h,w\rangle=(cw+d)^{-2h}(\bar c\bar w+\bar d)^{-2\bar h}|h,\bar h,\Lambda w\rangle,\\&\label{spin2}\phi_{h,\bar h}(\Lambda\cdot x;\Lambda w)=(cw+d)^{2h}(\bar c\bar w+\bar d)^{2\bar h}D_{h,\bar h}(\Lambda)\phi_{h,\bar h}(x;w),
\end{align}
where $D_{h,\bar h}(\Lambda)$ is a representation of the Lorentz transformation acting on the internal indices of  $\phi_{h,\bar h}(x,w)$.
\subsection{Generators}
\label{Sec3.3 }
In this last subsection we derive an explicit form of the generators in the two bases discussed above. We assume that a Lorentz transformation is parametrized by the variable $\alpha$, that can be either an angle $\varphi$ in the case of spatial rotations or a rapidity $\chi$ in the case of boosts. A spatial rotation about an axis $\hat{n}=(\sin\theta\cos\phi,\sin\theta\sin\phi,\cos\theta)$ of an angle $\varphi$ is given by the SL(2,$\mathbb C$) matrix $R_{\hat{n}}(\varphi)$
\begin{align}
    \label{gen3}
&R_{\hat{n}}(\varphi)=\left(\begin{matrix}\cos\frac{\varphi}{2}-i\cos\theta\sin\frac{\varphi}{2}& -i\sin\theta\sin\frac{\varphi}{2}e^{-i\phi}\\ \\-i\sin\theta\sin\frac{\varphi}{2}e^{i\phi} &\cos\frac{\varphi}{2}+i\cos\theta\sin\frac{\varphi}{2}
\end{matrix} \right), 
\end{align}
whereas a boost about $\hat{n}$ of rapidity $\chi$ by $B_{\hat{n}}(\chi)$
\begin{align}
\label{gen6}
B_{\hat{n}}(\chi)= \left(\begin{matrix}\cosh\frac{\chi}{2}-\cos\theta\sinh\frac{\chi}{2}& -\sin\theta\sinh\frac{\chi}{2}e^{-i\phi}\\ \\-\sin\theta\sinh\frac{\chi}{2}e^{i\phi} &\cosh\frac{\chi}{2}+\cos\theta\sinh\frac{\chi}{2}\end{matrix} \right).
\end{align}
The generators in the $(y,z,\bar z)$ basis can be easily computed from \eqref{coordtrans} as \cite{Alessio:2017lps},
\begin{align}
&G_z(\theta,\phi)=\frac{\partial}{\partial\alpha}\bigg(\frac{(a(\alpha)z+b(\alpha))(\bar{c}(\alpha)\bar{z}+\bar{d}(\alpha))+a(\alpha)\bar{c}(\alpha)y^2}{(c(\alpha)z+d(\alpha))(\bar{c}(\alpha)\bar{z}+\bar{d}(\alpha))+c(\alpha)\bar{c}(\alpha)y^2}\bigg)\bigg|_{\alpha=0},\qquad G_{\bar z}(\theta,\phi)=G^*_{ z}(\theta,\phi),\\
&G_y(\theta,\phi)=\frac{\partial}{\partial\alpha}\bigg(\frac{y}{(c(\alpha)z+d(\alpha))(\bar{c}(\alpha)\bar{z}+\bar{d}(\alpha))+c(\alpha)\bar{c}(\alpha)y^2}\bigg)\bigg|_{\alpha=0}.
\end{align}
and they are 
\begin{align}
 G(\theta,\phi)=G_z(\theta,\phi)\partial_z+ G_{\bar z}(\theta,\phi)\partial_{\bar z}+G_{y}(\theta,\phi)\partial_{y},
\end{align}
for each direction specified by the angles $(\theta,\phi)$. Explicitly they are\footnote{Note that in our conventions we use the inverse of the matrices in \eqref{gen3} and \eqref{gen6} corresponding to the inverse M\"obius transformation $w=\frac{d w'-b}{-c w'+a}$.}
\begin{align}
\label{G1}
&R_1(y,z)=\frac{i}{2}\bigg[(\bar z^2+y^2-1)\partial_{\bar z}-(z^2+y^2-1)\partial_z+y(\bar z-z)\partial_y\bigg],\\&R_2(y,z)=\frac{1}{2}\bigg[(\bar z^2-y^2+1)\partial_{\bar z}+(z^2-y^2+1)\partial_z+y(\bar z+z)\partial_y\bigg]\\& R_3(y,z)=-i(\bar z\partial_{\bar z}-z\partial_z),\\
&B_1(y,z)=-\frac{1}{2}\bigg[(\bar{z}^2-y^2-1)\partial_{\bar z}+(z^2-y^2-1)\partial_{z}+y(\bar z+z)\partial_y\bigg],
\\&B_2(y,z)=\frac{i}{2}\bigg[( \bar{z}^2+y^2+1)\partial_{\bar z}-(z^2+y^2+1)\partial_{z}+y(\bar z- z)\partial_y\bigg],\\\label{G3}
&B_3(y,z)=\bar{z}\partial_{\bar z}+z\partial_z+y\partial_y.
\end{align}
They satisfy the Lorentz algebra $[R_i(y,z),R_j(y,z)]=\epsilon^{ij}{}^{k}R_k(y,z)$, $[B_i(y,z),R_j(y,z)]=\epsilon^{ij}{}^{k}B_k(y,z)$ and $[B_i(y,z),B_j(y,z)]=-\epsilon^{ij}{}^{k}R_k(y,z)$.

Similarly, we can derive the generators in the basis $(\Delta,w,\bar w)$. Together with the coordinate part, that can be derived from \eqref{mobius}, now the generators comprise also an additional internal piece coming from \eqref{gen1.1}. We get
\begin{align}
    \label{gen2}
   G(\theta,\phi)= \frac{\partial}{\partial \alpha}|c(\alpha)w+d(\alpha)|^{2\Delta}   \bigg|_{\alpha=0}+\frac{\partial w}{\partial \alpha}\bigg|_{\alpha=0}\partial_{w}+\frac{\partial \bar{w}}{\partial \alpha}\bigg|_{\alpha=0}\partial_{\bar{w}},\qquad w=\frac{d w'-b}{-c w'+a}.
\end{align}
Explicitly \cite{Law:2019glh,Stieberger:2018onx},
\begin{align}
\label{gen5}
&R_1(\Delta,w)=\frac{i}{2}\bigg[(\bar w^2-1)\partial_{\bar w}-(w^2-1)\partial_{w}-\Delta( \bar w- w)\bigg],\\
&R_2(\Delta,w)=\frac{1}{2}\bigg[(\bar w^2+1)\partial_{\bar w}+(w^2+1)\partial_w-\Delta(w+\bar w)\bigg],\\
&R_3(\Delta,w)=-i(\bar w\partial_{\bar w}-w\partial_w),\\
&B_1(\Delta,w)=-\frac{1}{2}\bigg[(\bar{w}^2-1)\partial_{\bar w}+(w^2-1)\partial_{w}-\Delta(w+\bar w)\bigg],
\\&B_2(\Delta,w)=\frac{i}{2}\bigg[(\bar w^2+1)\partial_{\bar w}-(w^2+1)\partial_{w}-\Delta(\bar w-w)\bigg],\\
\label{B3}
&B_3(\Delta,w)=\bar{w}\partial_{\bar w}+w\partial_w-\Delta.
\end{align}
It is easy to show that the above generators satisfy the Lorentz algebra.
We notice that the differential part of the above generators can be obtained from the ones in \eqref{G1}-\eqref{G3} by sending $y\rightarrow 0$. Notice that, as emphasised at the beginning of this section, the conformal dimension $\Delta$ is the eigenvalue of boosts along the $x^3$ direction. This shows that the conformal primary basis just introduced diagonalizes the action of boosts along a given direction. 

In the more general case of spinning wavefunctions transforming as in \eqref{spin1} and \eqref{spin2}, the generators have to be modified in order to include the spin contribution. One can conclude by a similar computation that
\begin{align}
\label{F24}
&R_1(\Delta,w)=\frac{i}{2}\bigg[(\bar w^2-1)\partial_{\bar w}-(w^2-1)\partial_{w}-2( \bar h\bar w- hw)\bigg],\\
&R_2(\Delta,w)=\frac{1}{2}\bigg[(\bar w^2+1)\partial_{\bar w}+(w^2+1)\partial_w-2(hw+\bar h \bar w)\bigg],\\
&R_3(\Delta,w)=-i\bigg[\bar w\partial_{\bar w}-w\partial_w-(\bar h-h)\bigg],\\&B_1(\Delta,w)=-\frac{1}{2}\bigg[(\bar{w}^2-1)\partial_{\bar w}+(w^2-1)\partial_{w}-2(hw+\bar h\bar w)\bigg],
\\&B_2(\Delta,w)=\frac{i}{2}\bigg[(\bar w^2+1)\partial_{\bar w}-(w^2+1)\partial_{w}-2(\bar h\bar w-hw)\bigg],\\
&B_3(\Delta,w)=\bar{w}\partial_{\bar w}+w\partial_w-(h+\bar h)\,.
\end{align}
In this case it is clear that the generator of rotations along the $x^3$ direction contains an intrinsic extra-piece which is exactly the spin of the representation, given by $\sigma=h- \bar h$.

\section{Massive pairwise little group}
\label{Sec4}
So far we have focused on single-particle states introducing a basis which is diagonal under the action of boosts along a given direction. Such boosts are naturally a subgroup of the little group associated to this basis. The results of \cite{Gralla:2021eoi, Gralla:2021qaf} suggest the existence of a new quantum number associated to {\it pairs of particles} which, as discussed at the beginning of Section 3, renders the description of multiparticle states as ordinary tensor products of one-particle states inadequate. In this section we demonstrate that the new pairwise boost quantum number can be obtained from the little group of two-particle states in the conformal primary basis extended to include boost transformations along the direction of motion.

As discussed in \cite{Lippstreu:2021avq} in the conformal primary basis the little group of the state describing two massless particles in a frame where their spatial momenta are aligned, besides rotations around the axis of motion\footnote{As remarked in \cite{Zwanziger:1972sx} rotations around the common axis of motion obviously preserve both particles momenta.}, contains also boosts along the direction of motion, which we take again to be along the $x^3$ axis. Therefore, the little group of the two-particle state is generated by the pair $\{R_3,B_3\}$ satisfying $[R_3,B_3]=0$. Two massless momenta aligned along $x^3$ are described by stereographic coordinates $w_1=0$ and $w_2=\infty$ because, as remarked in Section \ref{Lor}, the angular coordinates of massless momenta are the same for bulk and boundary states. It follows that such reference two-particle states must be eigenstates of these operators with eigenvalues $\sigma_{12}$ and $\Delta_{12}$ and we denote them by $|\Delta_1,0;\Delta_2,\infty;\Delta_{12},\sigma_{12}\rangle$. Hence we have \cite{Banerjee:2018gce}
\begin{align}
    \label{mps}
&R_3(\Delta_1,w_1,\Delta_2,w_2,\Delta_{12})|\Delta_1,0;\Delta_2,\infty;\Delta_{12},\sigma_{12}\rangle= -i(\eta_1+\eta_2)\frac{\sigma_{12}}{2}|\Delta_1,0;\Delta_2,\infty;\Delta_{12},\sigma_{12}\rangle,\\&B_3(\Delta_1,w_1,\Delta_2,w_2,\Delta_{12})|\Delta_1,0;\Delta_2,\infty;\Delta_{12},\sigma_{12}\rangle=-(\eta_1+\eta_2)\frac{\Delta_{12}}{2}|\Delta_1,0;\Delta_2,\infty;\Delta_{12},\sigma_{12}\rangle,
\end{align}
and finite pairwise little group transformations with Wigner parameters $\alpha$ and $\beta$ read
\begin{align}
&U(R_3)|\Delta_1,0;\Delta_2,\infty;\Delta_{12},\sigma_{12}\rangle=
e^{\mp i \alpha\sigma_{12}}|\Delta_1,0;\Delta_2,\infty;\Delta_{12},\sigma_{12}\rangle,\\
&U(B_3)|\Delta_1,0;\Delta_2,\infty;\Delta_{12},\sigma_{12}\rangle=
e^{\mp \beta\Delta_{12}}|\Delta_1,0;\Delta_2,\infty;\Delta_{12},\sigma_{12}\rangle.
\end{align}
The eigenvalues $\sigma_{12}$ and $\Delta_{12}$ are referred to as {\it pairwise helicity} and {\it pairwise boost}. In the remainder of this paper we will only focus on $\Delta_{12}$ and its connection to the asymptotic boost-like angular momentum introduced in section \ref{sec1}. The explicit dependence of the Wigner parameter $\beta$ on the Lorentz transformation $\Lambda$ has been discussed in \cite{Lippstreu:2021avq} and it is
\begin{align}
\label{Lipp}
U(\Lambda)|\Delta_1,w_1;\Delta_2,w_2;\Delta_{12}\rangle=\prod_{i=1,2}|cw_i+d_i|^{-2\Delta_i+\eta_i\Delta_{12}}|\Delta_1,\Lambda w_1;\Delta_2,\Lambda w_2;\Delta_{12}\rangle.
\end{align}
Notice that the above formula, although derived for massless particles, holds also for massive particles. This is a simple consequence of the fact the in both the massive and massless case, the variables $w$ transform as in \eqref{mobius} under Lorentz transformations. 

Equation \eqref{Lipp} implies that the Lorentz generators acting on two-particle incoming and outgoing states in the conformal primary basis derived in \eqref{gen5}-\eqref{B3}, have to be modified according to the simple rule $\Delta_1\rightarrow\Delta_1-\frac{1}{2}\eta_2\Delta_{12}$ and $\Delta_2\rightarrow\Delta_2-\frac{1}{2}\eta_1\Delta_{12}$. In formulae,
\begin{align}
\label{gen12}
&R_1(\Delta_1,w_1;\Delta_2,w_2;\Delta_{12})=R_1(\Delta_1,w_1)+R_1(\Delta_2,w_2)+i\frac{\Delta_{12}}{4}[(\bar{w}_1-w_1)\eta_1+(\bar{w}_2-w_2)\eta_2],\\
&R_2(\Delta_1,w_1;\Delta_2,w_2;\Delta_{12})=R_2(\Delta_1,w_1)+R_2(\Delta_2,w_2)+\frac{\Delta_{12}}{4}[(\bar{w}_1+w_1)\eta_2+(\bar{w}_2+w_2)\eta_1],\\
&R_3(\Delta_1,w_1;\Delta_2,w_2;\Delta_{12})=R_3(\Delta_1,w_1)+R_3(\Delta_2,w_2),\\
&B_1(\Delta_1,w_1;\Delta_2,w_2;\Delta_{12})=B_1(\Delta_1,w_1)+B_1(\Delta_2,w_2)-\frac{\Delta_{12}}{4}[(w_1+\bar w_1)\eta_2+(w_2+\bar w_2)\eta_1],
\\&B_2(\Delta_1,w_1;\Delta_2,w_2;\Delta_{12})=B_2(\Delta_1,w_1)+B_2(\Delta_2,w_2)+i\frac{\Delta_{12}}{4}[(\bar w_1-w_1)\eta_2+(\bar w_2-w_2)\eta_1],\\
\label{GEN6}
&B_3(\Delta_1,w_1;\Delta_2,w_2;\Delta_{12})=B_3(\Delta_1,w_1)+B_3(\Delta_2,w_2)+\frac{\Delta_{12}}{2}(\eta_1+\eta_2).
\end{align}  
The expression for $B_3(\Delta_1,w_1;\Delta_2,w_2;\Delta_{12})$ above shows that two-particle states in the conformal primary basis carry a pairwise boost-like quantum number. From the signs of $\eta_1$ and $\eta_2$ we see that in a scattering process the particles will exchange an overall boost charge $
    \Delta B_3 =  2 \Delta_{12}$.
Such exchange is the analogue of the scoot effect discussed in Section \ref{sec1} with the crucial difference that in this case the states of the particles are described in the conformal primary basis while the scoot effect was derived using the conventional momentum basis. In what follows we demonstrate that the effect we just described is also present in the ordinary four-momentum parametrization of two-particle states thus providing a group-theoretic derivation of the scoot effect. Our analysis also extends the argument provided by Zwanziger \cite{Zwanziger:1972sx} for the existence of a pairwise helicity appearing in the scattering of electrically and magnetically charged particles to the boost-like pairwise charge. Before proceeding further let us stress here an important point. The generators in \eqref{gen12}-\eqref{GEN6} satisfy the Lorentz algebra even if $\Delta_{12}$ is not a constant, but a function of a Lorentz invariant quantity such as $q_1(w_1)\cdot q_2(w_2)$, where $q_i(w_i)$ are null momenta as in \eqref{nullmomx}. For simplicity, we will derive  the algebra in the on-shell momentum basis assuming that $\Delta_{12}$ is a constant. \\

Let us start by acting on \eqref{momb} with a Lorentz transformation
\begin{align}
\label{sp}
\nonumber U(\Lambda)|y,z\rangle &=\int_0^{\infty}d\lambda\int\hspace{0.05cm}dwd\bar w\hspace{0.05cm}G^*_{\Delta}(\hat{p};w)U(\Lambda)|\Delta,w\rangle\\&\stackrel{\eqref{F21}}{=}\int_0^{\infty}d\lambda\int\hspace{0.05cm}dwd\bar w\hspace{0.05cm}G^*_{\Delta}(\hat{p};w)|cw+d|^{-2\Delta}|\Delta,\Lambda w\rangle.
    \end{align}
Using now \eqref{cfb} one gets
    \begin{align}
        \label{sp2}
         \nonumber U(\Lambda)|y,z\rangle &=\int_0^{\infty}d\lambda\int\hspace{0.05cm}dwd\bar w\hspace{0.05cm}\int d\mu(p')G^*_{\Delta}(\hat{p};w)G_{\Delta}(\hat{p}';\Lambda w)|cw+d|^{-2\Delta}|y',z'\rangle,\\\nonumber &=\int_0^{\infty}d\lambda\int\hspace{0.05cm}dwd\bar w\hspace{0.05cm}\int d\mu(k)G^*_{\Delta}(\hat{p};w)G_{\Delta}(\hat{k}; w)|cw+d|^{-2\Delta}|cw+d|^{2\Delta}|\Lambda Y,\Lambda Z\rangle\\&=|\Lambda y,\Lambda z\rangle.
    \end{align}
 In going from the first to the second line we defined $\hat{p}'=\Lambda \hat{k}$ with $\hat{k}=\hat{k}(Y,Z)$ and then used Lorentz invariance of the measure. Equation \eqref{sp2} is just the standard result for the Lorentz transformations of scalar one-particle states. 
 
In order to find the form of the generators acting on two-particle states having a non-trivial pairwise boost charge $\Delta_{12}$
we need to compute the action of $U(\Lambda)$ on them. We denote this basis by $|p_1;p_2;\Delta_{12}\rangle\equiv|y_1,z_1;y_2,z_2;\Delta_{12}\rangle$. Repeating the same steps illustrated for single-particle states but now using \eqref{Lipp} we get
    \begin{align}
        \label{sp3}
        \nonumber &U(\Lambda)|y_1,z_1;y_2,z_2;\Delta_{12}\rangle=\int d\mu(\lambda_1,w_1)d\mu(\lambda_2,w_2)\int d\mu(k_1)d\mu(k_2)G_{\Delta_1}^*(\hat{p}_1;w_1)G_{\Delta_1}(\hat{k}_1;w_1)\\&\times G_{\Delta_2}^*(\hat{p}_2;w_2)G_{\Delta_2}(\hat{k}_2;w_2)|cw_1+d|^{-\eta_1\frac{\Delta_{12}}{2}}|cw_2+d|^{-\eta_2\frac{\Delta_{12}}{2}}|\Lambda Y_1,\Lambda Z_1;\Lambda Y_2,\Lambda Z_2;\Delta_{12}\rangle,
    \end{align}
   where $\hat{k}_i=\hat{k}_i(Y_i,Z_i)$  that we rewrite as
    \begin{align}
        \label{sp4}
        U(\Lambda)|y_1,z_1;y_2,z_2;\Delta_{12}\rangle=\int d\mu(k_1)d\mu(k_2) I_{\Delta_{12}}(\hat{p}_1;\hat{k}_1)I_{\Delta_{12}}(\hat{p}_2;\hat{k}_2)|\Lambda Y_1,\Lambda Z_1;\Lambda Y_2,\Lambda Z_2;\Delta_{12}\rangle,
    \end{align}
   with,
   \begin{align}
       \label{sp5}
       I_{\Delta_{12}}(\hat{p}_i;\hat{k}_i)=\int_0^{\infty}d\lambda_i\int\hspace{0.05cm}dw_id\bar w_iG_{\Delta_i}^*(\hat{p}_i;w_i)G_{\Delta_i}(\hat{k}_i;w_i)|cw_i+d|^{-\eta_i\frac{\Delta_{12}}{2}}\,.
   \end{align}
   Clearly for $\Delta_{12}=0$, using \eqref{ort3} we  have $I_{0}(\hat{p}_i,\hat{k}_i)=\delta^{(3)}(\hat{p}_i-\hat{k}_i)$. For $\Delta_{12}\neq 0$, as shown in Appendix \ref{appB}, we have 
\begin{align}
\nonumber &I_{\Delta_{12}}(\hat{p}_i,\hat{k}_i)  - I_{0}(\hat{p}_i,\hat{k}_i)\simeq -\eta_i\frac{\Delta_{12}}{2}(\alpha+\beta z_i+\bar\beta\bar z_i)I_0(\hat{p}_i,\hat{k}_i)\\\nonumber &-\eta_{i}\frac{\Delta_{12}}{4\pi}\int_0^{\infty} d\lambda_i\frac{y_i^{\Delta_i}Y^{\Delta^*_i}_i}{|\Gamma(i\lambda_i)|^2}\int_0^{\infty} dA A^{\Delta_i-3}e^{-A y^2_i}\int_0^{\infty} dB B^{\Delta^*_i-2}e^{-B Y^2_i}\\&\label{intzero}\int \frac{d^2\vec{l}}{(2\pi)^2}e^{i\vec{l}\cdot (\vec{Z}_i-\vec{z}_i)}e^{-\frac{l^2}{4A}}e^{-\frac{l^2}{4B}}[i(\beta+\bar\beta)l^1-(\beta-\bar\beta)l^2].
\end{align}
with $\alpha=d'(0)+\bar{d}'(0)$ and $\beta= c'(0).$
 We now show that the integral in the last two lines of the previous equation does not give any contribution to \eqref{sp4}. Indeed, we have to insert the previous equation in \eqref{sp4} and integrate it over $\hat{k}_i(Y_i,Z_i)$ obtaining the integral
\begin{align}
\int d^2\vec Z_ie^{i\vec{l}\cdot Z_i}=(2\pi)^2 \delta^{(2)}(\vec l_i).
\end{align}
Hence we have to set everywhere in the integral in the second and third line of \eqref{intzero} $\vec{l}=0$ and it gives exactly $0$ because it is linear in $\vec{l}$.
We have then
\begin{align}
    &[U(\Lambda)-\mathbf{1}]|y_1,z_1;y_2,z_2;\Delta_{12}\rangle\simeq\\& -\frac{\Delta_{12}}{2}[\alpha(\eta_1+\eta_2)+\beta (z_1\eta_1+z_2\eta_2)+\bar\beta(\bar z_1\eta_1+\bar z_2\eta_2)]|y_1,z_1;y_2,z_2;\Delta_{12}\rangle,
\end{align}
from which we can read the pairwise contribution to the generators in \eqref{G1}-\eqref{G3}. We have
\begin{align}\label{LGF1}
&R_1(y_1,z_1;y_2,w_2;\Delta_{12})=R_1(y_1,z_1)+R_1(y_2,z_2)+i\frac{\Delta_{12}}{4}[(\bar{z}_1-z_1)\eta_1+(\bar{z}_2-z_2)\eta_2],\\
&R_2(y_1,z_1;y_2,w_2;\Delta_{12})=R_2(y_1,z_1)+R_2(y_2,z_2)+\frac{\Delta_{12}}{4}[(\bar{z}_1+z_1)\eta_2+(\bar{z}_2+z_2)\eta_1],\\
&R_3(y_1,z_1;y_2,w_2;\Delta_{12})=R_3(y_1,z_1)+R_3(y_2,z_2),\\
&B_1(y_1,z_1;y_2,w_2;\Delta_{12})=B_1(y_1,z_1)+B_1(y_2,z_2)-\frac{\Delta_{12}}{4}[(z_1+\bar z_1)\eta_2+(z_2+\bar z_2)\eta_1],
\\&B_2(y_1,z_1;y_2,w_2;\Delta_{12})=B_2(y_1,z_1)+B_2(y_2,z_2)+i\frac{\Delta_{12}}{4}[(\bar z_1-z_1)\eta_2+(\bar z_2-z_2)\eta_1],\\
&B_3(y_1,z_1;y_2,w_2;\Delta_{12})=B_3(y_1,z_1)+B_3(y_2,z_2)+\frac{\Delta_{12}}{2}(\eta_1+\eta_2),\label{LGF2}
\end{align}
In the center of mass frame, where $z_i=\bar z_i=0$, we have
\begin{align}
\Delta B_3=B_3^{\mathrm{out}}-B_3^{\mathrm{in}}=2\Delta_{12}.
\end{align}
Comparing with equation \eqref{match}, we can make the identification
\begin{align}
\Delta_{12}=-\Delta_{21}=\frac{e_1e_2}{\gamma^2-1}\log\bigg|\frac{\tau_1}{\tau_2}\bigg|,
\end{align}
in electromagnetism and
\begin{align}
\Delta_{12}=-\Delta_{21}=\frac{2Gm_1m_2\gamma(3-2\gamma^2)}{\gamma^2-1}\log\bigg|\frac{\tau_1}{\tau_2}\bigg|,
\end{align}
in gravity, where the Lorentz factor $\gamma$ is given by 
\begin{align}
\gamma=-\hat{p}_1(y_1,z_1)\cdot\hat{p}_2(y_2,z_2)=\frac{y_1^2+y_2^2+|z_{12}|^2}{y_1y_2},\qquad z_{12}=z_1-z_2\,.
\end{align}
Notice that, with this identification\footnote{One can show that, keeping a non-vanishing pairwise helicity $\sigma_{12}$ and repeating the matching procedure with the asymptotic angular momentum found in \cite{Zwanziger:1972sx}, it will imply simply $\sigma_{12}\sim\mu_{12}=e_1g_2-e_2g_1$ without additional dependence on the particles four-momenta.}, $\Delta_{12}$ depends on the two on-shell momenta through the Lorentz factor $\gamma$. It is not immediately obvious that in this case the modified generators in \eqref{LGF1}-\eqref{LGF2} will still satisfy the Lorentz algebra. However, it can be show that this is the case i.e. $[R_i(y_i,z_i;\Delta_{12}),R_j(y_i,z_i;\Delta_{12})]=\epsilon^{ij}{}^{k}R_k(y_i,z_i;\Delta_{12})$, $[B_i(y_i,z_i;\Delta_{12}),R_j(y_i,z_i;\Delta_{12})]=\epsilon^{ij}{}^{k}B_k(y_i,z_i;\Delta_{12})$ and $[B_i(y_i,z_i;\Delta_{12}),B_j(y_i,z_i;\Delta_{12})]=-\epsilon^{ij}{}^{k}R_k(y_i,z_i;\Delta_{12})$ since $\gamma$ is a Lorentz invariant quantity. 

This completes our proof that the asymptotic boost charge carried by multi-particle states in scattering processes in electromagnetism and gravity can be obtained in a group-theoretic fashion as a pairwise quantum number linked to the little group of celestial multi-particle states.

\section{Conclusions}

In his original 1972 paper \cite{Zwanziger:1972sx}, Zwanziger observed that the asymptotic electromagnetic field generated by pairs of particles with both electric and magnetic charges carries a non-vanishing angular momentum due to the charge-pole interaction. The group-theoretic origin of this new quantum number was identified in a modification of the action of Lorentz generators on scattering states associated to the little group of the four-momenta associated to pairs of particles.

Zwanziger also identified potential pairwise Coulombic contributions (depending only on the product of the particles electric and magnetic charges separately) to the asymptotic relativistic angular momentum of the electromagnetic field but dismissed them as vanishing or ill-defined. In retrospect this conclusion appears also to be justified by the fact that, in the ordinary four-momentum basis for scattering states, the only manifest transformations which leave invariant the four-momentum configuration of pairs of particles are rotations around a given spatial axis.

Recent results \cite{Gralla:2021eoi} motivated by gravitational scattering calculations suggest, however, that already in ordinary electromagnetism, in the absence of magnetic charges, the electromagnetic field possesses an asymptotic boost-like angular momentum proving that the Coulombic contributions originally negelcted by Zwanziger are, in fact, real. Indeed what has been dubbed as ``scoot effect" consists in an exchange of boost angular momentum between particles and field. A natural question then concerns the group theoretic-origin of such boost-like angular momentum carried by asymptotic states. In this work we provided an answer to this question showing that the origin of the pairwise boost charge can be traced back to the little group of a two-particle state {\it in the conformal primary basis representation}. We explicitly showed that, using the integral transform connecting the ordinary on-shell momentum representation to the conformal primary one, the pairwise boost charge contribution to two-particle states in the conformal primary basis translates into an analogous charge for two-particle states in the ordinary momentum representation. Such charge appears in a modification of the expression of the generators of Lorentz transformations acting on two-particle states \eqref{LGF1}-\eqref{LGF2}. In the case of the pairwise helicity discovered by Zwanziger this modified action of Lorentz generators on multiparticle states has been interpreted \cite{Csaki:2020yei,Csaki:2020inw,Lippstreu:2021avq} in terms of an extension of the Fock space representation of the Poincaré group in which two-particle states are no longer described by simple tensor products of one-particle states but contain an additional factor representing a ``pairwise Hilbert space" in which the pairwise helicity is stored. Alternatively one could keep the ordinary tensor product structure of the two-particle states and interpret the modification of the action of Lorentz generators as a modification of the ordinary Leibniz action on tensor product states carrying an additional term accounting for the pairwise angular momentum. Either way the existence of a pairwise relativistic angular momentum carried by asymptotic states renders the conventional Fock space picture of multiparticle states and their associated observables inadequate. In this work we pointed out that, besides electromagnetic scattering, long range effects in gravitational scattering require the existence of a pairwise boost-like angular momentum and we evidenced the subtle connection between this quantum number and the little group of of two-particle states in the conformal primary basis. A remarkable upshot of our analysis is that, given the universal character of the coupling of gravity to all forms of matter and energy, the inadequacy of a Fock space description of multiparticle states appears to be a general infrared consequence of the existence of gravitational interactions. Given that the Fock space structure of multiparticle states in free quantum field theory is a cornerstone of the description of particle creation effects in the presence of causal horizons, it is tempting to speculate that an infrared failure of such framework might have significant consequences and challenge our current understanding of puzzling aspects like, for example, the apparent failure of unitarity in black hole quantum evolution. We leave further exploration of these implications to future studies.
\subsection*{Acknowledgements}
The research of FA is fully supported by the Knut and Alice Wallenberg Foundation under grant KAW 2018.0116. MA acknowledges support from the INFN Iniziativa Specifica QUAGRAP. The research of MA was also carried out in the frame of Programme STAR Plus, financially supported by the University of Napoli Federico II and Compagnia di San Paolo.
\appendix
\section{Transformation of massive momenta}
\label{appA}
The isomorphism between the Lorentz group and SL(2,$\mathbb{C}/\mathbb{Z}^2$) is realized as \cite{Oblak:2015qia}

\begin{eqnarray}
\nonumber &\Lambda^\mu{}_\nu(a,b,c,d) =\\\nonumber &=\frac{1}{2}\begin{pmatrix} a\bar{a}+b\bar{b}+c\bar{c}+d\bar{d} & b\bar{a}+a\bar{b}+d\bar{c}+c\bar{d} & i(-b\bar{a}+a\bar{b}-d\bar{c}+c\bar{d}) & -a\bar{a}+b\bar{b}-c\bar{c}+d\bar{d}\cr c\bar{a}+a\bar{c}+d\bar{b}+b\bar{d} & d\bar{a}+a\bar{d}+c\bar{b}+b\bar{c} & i(-d\bar{a}+a\bar{d}+c\bar{b}-b\bar{c}) & -c\bar{a}-a\bar{c}+d\bar{b}+b\bar{d}\cr i(c\bar{a}-a\bar{c}+d\bar{b}-b\bar{d}) & i(d\bar{a}-a\bar{d}+c\bar{b}-b\bar{c}) & d\bar{a}+a\bar{d}-c\bar{b}-b\bar{c} & i(-c\bar{a}+a\bar{c}+d\bar{b}-b\bar{d})\cr -a\bar{a}-b\bar{b}+c\bar{c}+d\bar{d} & -b\bar{a}-a\bar{b}+d\bar{c}+c\bar{d} & i(b\bar{a}-a\bar{b}-d\bar{c}+c\bar{d}) & a\bar{a}-b\bar{b}-c\bar{c}+d\bar{d}\end{pmatrix},
\end{eqnarray}
where $ad-bc=\bar a\bar d-\bar b\bar c=1$. When acting on massive momenta $\hat{p}^{\mu}$ in \eqref{momzy}
we get
\begin{align}
\nonumber&\hat{p}'^{\mu}(y',z')=\Lambda^{\mu}{}_{\nu}(a,b,c,d)\hat{p}^{\nu}(y,z)\\&\nonumber=\frac{1}{2y}(y^2(|a|^2+|c|^2)+|az+b|^2+|cz+d|^2,y^2(c \bar a+a\bar c)+(cz+d)(\bar a\bar z+\bar b)+(az+b)(\bar c\bar z+\bar d),\\\nonumber&i[y^2(\bar a c-a\bar c)-(az+b)(\bar c\bar z+\bar d)+(cz+d)(\bar a\bar z+\bar b)],y^2(|c|^2-|a|^2)-|az+b|^2+|cz+d|^2)\\&=\frac{1}
{2y'} \left(1+y'^2+z'\bar{z}' ,z'+\bar{z}',i (\bar{z}'-z'),1-y'^2-z'\bar{z}'\right),
\end{align}
with 
\begin{align}
z' = \frac{(az+b)(\bar{c}\bar{z}+\bar{d})+a\bar{c}y^2}{(cz+d)(\bar{c}\bar{z}+\bar{d})+c\bar{c}y^2}, \qquad  y' = \frac{y}{(cz+d)(\bar{c}\bar{z}+\bar{d})+c\bar{c}y^2},
\end{align}
as displayed in \eqref{coordtrans}.
\section{Pairwise integral}
\label{appB}
In this Appendix we analyse the integral appearing in \eqref{sp5}. Choosing the parametrization $\hat{p}_i=\hat{p}_i(Y_i,Z_i)$ and $\hat{k}_i=\hat{k}_i(y_i,z_i)$ we can re-write the integral as
   \begin{align}
        \label{sp6}
        \nonumber I_{\Delta_{12}}(\hat{p}_i,\hat{k}_i)=&\frac{1}{4\pi^3}\int_0^{\infty}d\lambda\int\hspace{0.05cm}dw d\bar w \frac{\Gamma(\Delta)\Gamma(\Delta^*)}{\Gamma(i\lambda)\Gamma(-i\lambda)}\bigg(\frac{y_i}{y^2_i+|w-z_i |^2}\bigg)^{\Delta}\bigg(\frac{Y_i}{Y_i+|w-Z_i|^2}\bigg)^{\Delta^*}\\&\times|cw+d|^{-\eta_{i}\frac{\Delta_{12}}{2}}. \end{align}
 We now consider the case with $\Delta_{12}\neq 0$. In order to find the generator we need to compute the derivative
\begin{align}
\frac{d}{d x}(|c(x)w+d(x)|^2)^{-\eta_{i}\frac{\Delta_{12}}{2}}|_{x=0}=-\eta_{i}\frac{\Delta_{12}}{2}(\alpha+\beta w+\bar\beta\bar w)=f(w,\bar w), 
\end{align}
where we used $d(0)=1$ and $c(0)=0$, and where $
\alpha\equiv d'(0)+\bar{d}'(0),\,\beta\equiv c'(0)$. Therefore, we see that the new integral we have to compute is
\begin{align}
 \nonumber &I_{\Delta_{12}}(\hat{p}_i,\hat{k}_i)  \simeq I_{0}(\hat{p}_i,\hat{k}_i)\\&+\frac{1}{4\pi^3}\int_0^{\infty}d\lambda\int\hspace{0.05cm}dw d\bar w \frac{\Gamma(\Delta)\Gamma(\Delta^*)}{\Gamma(i\lambda)\Gamma(-i\lambda)}\bigg(\frac{y_i}{y_i^2+|w-Z_i|^2}\bigg)^{\Delta}\bigg(\frac{Y_i}{Y^2_i+|w-Z_i|^2}\bigg)^{\Delta^*}f(w,\bar{w}).
\end{align}
We have, using $w=x^1+ix^2$ and $dwd\bar{w}=2d^2\vec{x}$ we get
\begin{align}
\nonumber &I_{\Delta_{12}}(\hat{p}_i,\hat{k}_i)  \simeq I_{0}(\hat{p}_i,\hat{k}_i)\\&+\frac{1}{2\pi^3}\int_0^{\infty}d\lambda\frac{|\Gamma(\Delta)|^2y_i^{\Delta}Y_i^{\Delta^*}}{|\Gamma(i\lambda)|^2}\int d^2\vec{x}\bigg(\frac{1}{y^2_i+|\vec{x}-\vec{z}_i|^2}\bigg)^{\Delta}\bigg(\frac{1}{Y^2_i+|\vec{x}-\vec{Z}_i|^2}\bigg)^{\Delta^*}f(\vec x),
\end{align}
that, using Schwinger parametrization can be rewritten as
\begin{align}
   \nonumber &I_{\Delta_{12}}(\hat{p}_i,\hat{k}_i)  - I_{0}(\hat{p}_i,\hat{k}_i)\\\nonumber &\simeq \frac{1}{2\pi^3}\int_0^{\infty} d\lambda\frac{y_i^{\Delta}Y_i^{\Delta^*}}{|\Gamma(i\lambda)|^2}\int \frac{d^2\vec{l}}{(2\pi)^2}e^{i\vec{l}\cdot (\vec{x}-\vec{\rho})}\\&\times \int_0^{\infty} dA A^{\Delta-1}e^{-A y^2_i}\int_0^{\infty} dB B^{\Delta^*-1}e^{-B Y^2_i}\int d^2\vec{x}e^{-A|\vec{x}-\vec{z}_i|^2}f(\vec{x})\int d^2\vec{\rho}e^{-B|\vec{\rho}-\vec{Z}_i|^2}.
\end{align}
We perform the shifts $\vec{x}'=\vec{x}-\vec{z}_i$ and $\vec{\rho}'=\vec{\rho}-\vec{Z}_i$ and we get
\begin{align}
    \nonumber &I_{\Delta_{12}}(\hat{p}_i,\hat{k}_i)  - I_{0}(\hat{p}_i,\hat{k}_i)\\\nonumber &\simeq \frac{1}{2\pi^3}\int_0^{\infty} d\lambda\frac{y_i^{\Delta}Y_i^{\Delta^*}}{|\Gamma(i\lambda)|^2}\int \frac{d^2\vec{l}}{(2\pi)^2}e^{i\vec{l}\cdot (\vec{z}_i-\vec{Z}_i)}\int_0^{\infty} dA A^{\Delta-1}e^{-A y^2_i}e^{-\frac{l^2}{4A}}\int_0^{\infty} dB B^{\Delta^*-1}e^{-B Y^2_i}e^{-\frac{l^2}{4B}}\\\nonumber&\times \int d^2\vec{x}e^{-A(\vec{x}-i\frac{\vec{l}}{2A})^2}f(\vec{x}+\vec{z}_i)\int d^2\vec{\rho}e^{-B(\vec{\rho}-i\frac{\vec{l}}{2B})^2}\\\nonumber &= \frac{1}{2\pi^3}\int_0^{\infty} d\lambda\frac{y_i^{\Delta}Y_i^{\Delta^*}}{|\Gamma(i\lambda)|^2}\int \frac{d^2\vec{k}}{(2\pi)^2}e^{i\vec{l}\cdot (\vec{z}_i-\vec{Z}_i)}\int_0^{\infty} dA A^{\Delta-1}e^{-A y^2_i}e^{-\frac{l^2}{4A}}\int_0^{\infty} dB B^{\Delta^*-1}e^{-B Y^2_i}e^{-\frac{l^2}{4B}}\\&\times \int d^2\vec{x}e^{-A\vec{x}^2}f\bigg(\vec{x}+\vec{z}+i\frac{\vec{l}}{2A}\bigg)\int d^2\vec{\rho}e^{-B\vec{\rho}^2}.
\end{align}
We need to compute gaussian integral
\begin{align}
\nonumber &\int d^2\vec{x}e^{-A\vec{x}^2}\bigg[\alpha+\beta\bigg(x^1+ix^2+z^1_i+iz^2_i+i\frac{l^1+il^2}{2A}\bigg)+\bar\beta\bigg(x^1-ix^2+z^1_i-iz^2_i+i\frac{l^1-il^2}{2A}\bigg)\bigg]\\=&\frac{\pi}{A}\bigg[\alpha+(\beta+\bar\beta)\bigg(z^1_i+i\frac{l^1}{2A}\bigg)+i(\beta-\bar\beta)\bigg(z_i^2+i\frac{l^2}{2A}\bigg)\bigg],
\end{align}
that yields equation \eqref{intzero}.

\bibliography{bibliography}

\begin{thebibliography}{61}
\expandafter\ifx\csname natexlab\endcsname\relax\def\natexlab#1{#1}\fi
\expandafter\ifx\csname bibnamefont\endcsname\relax
  \def\bibnamefont#1{#1}\fi
\expandafter\ifx\csname bibfnamefont\endcsname\relax
  \def\bibfnamefont#1{#1}\fi
\expandafter\ifx\csname citenamefont\endcsname\relax
  \def\citenamefont#1{#1}\fi
\expandafter\ifx\csname url\endcsname\relax
  \def\url#1{\texttt{#1}}\fi
\expandafter\ifx\csname urlprefix\endcsname\relax\def\urlprefix{URL }\fi
\providecommand{\bibinfo}[2]{#2}
\providecommand{\eprint}[2][]{\url{#2}}

\bibitem[{\citenamefont{Zwanziger}(1972)}]{Zwanziger:1972sx}
\bibinfo{author}{\bibfnamefont{D.}~\bibnamefont{Zwanziger}},
  \bibinfo{journal}{Phys. Rev. D} \textbf{\bibinfo{volume}{6}},
  \bibinfo{pages}{458} (\bibinfo{year}{1972}).

\bibitem[{\citenamefont{Cs\'aki et~al.}(2021)\citenamefont{Cs\'aki, Hong,
  Shirman, Telem, and Terning}}]{Csaki:2020yei}
\bibinfo{author}{\bibfnamefont{C.}~\bibnamefont{Cs\'aki}},
  \bibinfo{author}{\bibfnamefont{S.}~\bibnamefont{Hong}},
  \bibinfo{author}{\bibfnamefont{Y.}~\bibnamefont{Shirman}},
  \bibinfo{author}{\bibfnamefont{O.}~\bibnamefont{Telem}}, \bibnamefont{and}
  \bibinfo{author}{\bibfnamefont{J.}~\bibnamefont{Terning}},
  \bibinfo{journal}{Phys. Rev. Lett.} \textbf{\bibinfo{volume}{127}},
  \bibinfo{pages}{041601} (\bibinfo{year}{2021}), \eprint{2010.13794}.

\bibitem[{\citenamefont{Cs\'aki et~al.}(2022)\citenamefont{Cs\'aki, Shirman,
  Telem, and Terning}}]{Csaki:2022qtz}
\bibinfo{author}{\bibfnamefont{C.}~\bibnamefont{Cs\'aki}},
  \bibinfo{author}{\bibfnamefont{Y.}~\bibnamefont{Shirman}},
  \bibinfo{author}{\bibfnamefont{O.}~\bibnamefont{Telem}}, \bibnamefont{and}
  \bibinfo{author}{\bibfnamefont{J.}~\bibnamefont{Terning}},
  \bibinfo{journal}{Phys. Rev. Lett.} \textbf{\bibinfo{volume}{129}},
  \bibinfo{pages}{181601} (\bibinfo{year}{2022}).

\bibitem[{\citenamefont{Csaki et~al.}(2021)\citenamefont{Csaki, Hong, Shirman,
  Telem, Terning, and Waterbury}}]{Csaki:2020inw}
\bibinfo{author}{\bibfnamefont{C.}~\bibnamefont{Csaki}},
  \bibinfo{author}{\bibfnamefont{S.}~\bibnamefont{Hong}},
  \bibinfo{author}{\bibfnamefont{Y.}~\bibnamefont{Shirman}},
  \bibinfo{author}{\bibfnamefont{O.}~\bibnamefont{Telem}},
  \bibinfo{author}{\bibfnamefont{J.}~\bibnamefont{Terning}}, \bibnamefont{and}
  \bibinfo{author}{\bibfnamefont{M.}~\bibnamefont{Waterbury}},
  \bibinfo{journal}{JHEP} \textbf{\bibinfo{volume}{08}}, \bibinfo{pages}{029}
  (\bibinfo{year}{2021}), \eprint{2009.14213}.

\bibitem[{\citenamefont{Gralla and Lobo}(2022{\natexlab{a}})}]{Gralla:2021eoi}
\bibinfo{author}{\bibfnamefont{S.~E.} \bibnamefont{Gralla}} \bibnamefont{and}
  \bibinfo{author}{\bibfnamefont{K.}~\bibnamefont{Lobo}},
  \bibinfo{journal}{Phys. Rev. D} \textbf{\bibinfo{volume}{105}},
  \bibinfo{pages}{084053} (\bibinfo{year}{2022}{\natexlab{a}}),
  \eprint{2112.01729}.

\bibitem[{\citenamefont{Bhardwaj and Lippstreu}(2022)}]{Bhardwaj:2022hip}
\bibinfo{author}{\bibfnamefont{R.}~\bibnamefont{Bhardwaj}} \bibnamefont{and}
  \bibinfo{author}{\bibfnamefont{L.}~\bibnamefont{Lippstreu}}
  (\bibinfo{year}{2022}), \eprint{2208.02727}.

\bibitem[{\citenamefont{Gralla and Lobo}(2022{\natexlab{b}})}]{Gralla:2021qaf}
\bibinfo{author}{\bibfnamefont{S.~E.} \bibnamefont{Gralla}} \bibnamefont{and}
  \bibinfo{author}{\bibfnamefont{K.}~\bibnamefont{Lobo}},
  \bibinfo{journal}{Class. Quant. Grav.} \textbf{\bibinfo{volume}{39}},
  \bibinfo{pages}{095001} (\bibinfo{year}{2022}{\natexlab{b}}),
  \eprint{2110.08681}.

\bibitem[{\citenamefont{Laddha and Sen}(2018)}]{Laddha:2018myi}
\bibinfo{author}{\bibfnamefont{A.}~\bibnamefont{Laddha}} \bibnamefont{and}
  \bibinfo{author}{\bibfnamefont{A.}~\bibnamefont{Sen}},
  \bibinfo{journal}{JHEP} \textbf{\bibinfo{volume}{10}}, \bibinfo{pages}{056}
  (\bibinfo{year}{2018}), \eprint{1804.09193}.

\bibitem[{\citenamefont{Sahoo and Sen}(2019)}]{Sahoo:2018lxl}
\bibinfo{author}{\bibfnamefont{B.}~\bibnamefont{Sahoo}} \bibnamefont{and}
  \bibinfo{author}{\bibfnamefont{A.}~\bibnamefont{Sen}},
  \bibinfo{journal}{JHEP} \textbf{\bibinfo{volume}{02}}, \bibinfo{pages}{086}
  (\bibinfo{year}{2019}), \eprint{1808.03288}.

\bibitem[{\citenamefont{Saha et~al.}(2020)\citenamefont{Saha, Sahoo, and
  Sen}}]{Saha:2019tub}
\bibinfo{author}{\bibfnamefont{A.~P.} \bibnamefont{Saha}},
  \bibinfo{author}{\bibfnamefont{B.}~\bibnamefont{Sahoo}}, \bibnamefont{and}
  \bibinfo{author}{\bibfnamefont{A.}~\bibnamefont{Sen}},
  \bibinfo{journal}{JHEP} \textbf{\bibinfo{volume}{06}}, \bibinfo{pages}{153}
  (\bibinfo{year}{2020}), \eprint{1912.06413}.

\bibitem[{\citenamefont{Sahoo}(2020)}]{Sahoo:2020ryf}
\bibinfo{author}{\bibfnamefont{B.}~\bibnamefont{Sahoo}},
  \bibinfo{journal}{JHEP} \textbf{\bibinfo{volume}{12}}, \bibinfo{pages}{070}
  (\bibinfo{year}{2020}), \eprint{2008.04376}.

\bibitem[{\citenamefont{Krishna and Sahoo}(2023)}]{Krishna:2023fxg}
\bibinfo{author}{\bibfnamefont{H.}~\bibnamefont{Krishna}} \bibnamefont{and}
  \bibinfo{author}{\bibfnamefont{B.}~\bibnamefont{Sahoo}},
  \bibinfo{journal}{JHEP} \textbf{\bibinfo{volume}{11}}, \bibinfo{pages}{233}
  (\bibinfo{year}{2023}), \eprint{2308.16807}.

\bibitem[{\citenamefont{Pasterski et~al.}(2017)\citenamefont{Pasterski, Shao,
  and Strominger}}]{Pasterski:2016qvg}
\bibinfo{author}{\bibfnamefont{S.}~\bibnamefont{Pasterski}},
  \bibinfo{author}{\bibfnamefont{S.-H.} \bibnamefont{Shao}}, \bibnamefont{and}
  \bibinfo{author}{\bibfnamefont{A.}~\bibnamefont{Strominger}},
  \bibinfo{journal}{Phys. Rev. D} \textbf{\bibinfo{volume}{96}},
  \bibinfo{pages}{065026} (\bibinfo{year}{2017}), \eprint{1701.00049}.

\bibitem[{\citenamefont{Pasterski and Shao}(2017)}]{Pasterski:2017kqt}
\bibinfo{author}{\bibfnamefont{S.}~\bibnamefont{Pasterski}} \bibnamefont{and}
  \bibinfo{author}{\bibfnamefont{S.-H.} \bibnamefont{Shao}},
  \bibinfo{journal}{Phys. Rev. D} \textbf{\bibinfo{volume}{96}},
  \bibinfo{pages}{065022} (\bibinfo{year}{2017}), \eprint{1705.01027}.

\bibitem[{\citenamefont{Pasterski}(2021)}]{Pasterski:2021rjz}
\bibinfo{author}{\bibfnamefont{S.}~\bibnamefont{Pasterski}},
  \bibinfo{journal}{Eur. Phys. J. C} \textbf{\bibinfo{volume}{81}},
  \bibinfo{pages}{1062} (\bibinfo{year}{2021}), \eprint{2108.04801}.

\bibitem[{\citenamefont{Pasterski et~al.}(2021)\citenamefont{Pasterski, Pate,
  and Raclariu}}]{Pasterski:2021raf}
\bibinfo{author}{\bibfnamefont{S.}~\bibnamefont{Pasterski}},
  \bibinfo{author}{\bibfnamefont{M.}~\bibnamefont{Pate}}, \bibnamefont{and}
  \bibinfo{author}{\bibfnamefont{A.-M.} \bibnamefont{Raclariu}}, in
  \emph{\bibinfo{booktitle}{{Snowmass 2021}}} (\bibinfo{year}{2021}),
  \eprint{2111.11392}.

\bibitem[{\citenamefont{Arkani-Hamed et~al.}(2021)\citenamefont{Arkani-Hamed,
  Pate, Raclariu, and Strominger}}]{Arkani-Hamed:2020gyp}
\bibinfo{author}{\bibfnamefont{N.}~\bibnamefont{Arkani-Hamed}},
  \bibinfo{author}{\bibfnamefont{M.}~\bibnamefont{Pate}},
  \bibinfo{author}{\bibfnamefont{A.-M.} \bibnamefont{Raclariu}},
  \bibnamefont{and}
  \bibinfo{author}{\bibfnamefont{A.}~\bibnamefont{Strominger}},
  \bibinfo{journal}{JHEP} \textbf{\bibinfo{volume}{08}}, \bibinfo{pages}{062}
  (\bibinfo{year}{2021}), \eprint{2012.04208}.

\bibitem[{\citenamefont{Raclariu}(2021)}]{Raclariu:2021zjz}
\bibinfo{author}{\bibfnamefont{A.-M.} \bibnamefont{Raclariu}}
  (\bibinfo{year}{2021}), \eprint{2107.02075}.

\bibitem[{\citenamefont{Iacobacci et~al.}(2024)\citenamefont{Iacobacci,
  Sleight, and Taronna}}]{Iacobacci:2024nhw}
\bibinfo{author}{\bibfnamefont{L.}~\bibnamefont{Iacobacci}},
  \bibinfo{author}{\bibfnamefont{C.}~\bibnamefont{Sleight}}, \bibnamefont{and}
  \bibinfo{author}{\bibfnamefont{M.}~\bibnamefont{Taronna}}
  (\bibinfo{year}{2024}), \eprint{2401.16591}.

\bibitem[{\citenamefont{Iacobacci and M\"uck}(2020)}]{Iacobacci:2020por}
\bibinfo{author}{\bibfnamefont{L.}~\bibnamefont{Iacobacci}} \bibnamefont{and}
  \bibinfo{author}{\bibfnamefont{W.}~\bibnamefont{M\"uck}},
  \bibinfo{journal}{Phys. Rev. D} \textbf{\bibinfo{volume}{102}},
  \bibinfo{pages}{106025} (\bibinfo{year}{2020}), \eprint{2009.02938}.

\bibitem[{\citenamefont{Sleight and Taronna}(2023)}]{Sleight:2023ojm}
\bibinfo{author}{\bibfnamefont{C.}~\bibnamefont{Sleight}} \bibnamefont{and}
  \bibinfo{author}{\bibfnamefont{M.}~\bibnamefont{Taronna}}
  (\bibinfo{year}{2023}), \eprint{2301.01810}.

\bibitem[{\citenamefont{Ball et~al.}(2019)\citenamefont{Ball, Himwich,
  Narayanan, Pasterski, and Strominger}}]{Ball:2019atb}
\bibinfo{author}{\bibfnamefont{A.}~\bibnamefont{Ball}},
  \bibinfo{author}{\bibfnamefont{E.}~\bibnamefont{Himwich}},
  \bibinfo{author}{\bibfnamefont{S.~A.} \bibnamefont{Narayanan}},
  \bibinfo{author}{\bibfnamefont{S.}~\bibnamefont{Pasterski}},
  \bibnamefont{and}
  \bibinfo{author}{\bibfnamefont{A.}~\bibnamefont{Strominger}},
  \bibinfo{journal}{JHEP} \textbf{\bibinfo{volume}{08}}, \bibinfo{pages}{168}
  (\bibinfo{year}{2019}), \eprint{1905.09809}.

\bibitem[{\citenamefont{Pate et~al.}(2021)\citenamefont{Pate, Raclariu,
  Strominger, and Yuan}}]{Pate:2019lpp}
\bibinfo{author}{\bibfnamefont{M.}~\bibnamefont{Pate}},
  \bibinfo{author}{\bibfnamefont{A.-M.} \bibnamefont{Raclariu}},
  \bibinfo{author}{\bibfnamefont{A.}~\bibnamefont{Strominger}},
  \bibnamefont{and} \bibinfo{author}{\bibfnamefont{E.~Y.} \bibnamefont{Yuan}},
  \bibinfo{journal}{Rev. Math. Phys.} \textbf{\bibinfo{volume}{33}},
  \bibinfo{pages}{2140003} (\bibinfo{year}{2021}), \eprint{1910.07424}.

\bibitem[{\citenamefont{Fotopoulos and Taylor}(2019)}]{Fotopoulos:2019tpe}
\bibinfo{author}{\bibfnamefont{A.}~\bibnamefont{Fotopoulos}} \bibnamefont{and}
  \bibinfo{author}{\bibfnamefont{T.~R.} \bibnamefont{Taylor}},
  \bibinfo{journal}{JHEP} \textbf{\bibinfo{volume}{10}}, \bibinfo{pages}{167}
  (\bibinfo{year}{2019}), \eprint{1906.10149}.

\bibitem[{\citenamefont{Law and Zlotnikov}(2020{\natexlab{a}})}]{Law:2020tsg}
\bibinfo{author}{\bibfnamefont{Y.~T.~A.} \bibnamefont{Law}} \bibnamefont{and}
  \bibinfo{author}{\bibfnamefont{M.}~\bibnamefont{Zlotnikov}},
  \bibinfo{journal}{JHEP} \textbf{\bibinfo{volume}{06}}, \bibinfo{pages}{079}
  (\bibinfo{year}{2020}{\natexlab{a}}), \eprint{2004.04309}.

\bibitem[{\citenamefont{Fotopoulos et~al.}(2020)\citenamefont{Fotopoulos,
  Stieberger, Taylor, and Zhu}}]{Fotopoulos:2020bqj}
\bibinfo{author}{\bibfnamefont{A.}~\bibnamefont{Fotopoulos}},
  \bibinfo{author}{\bibfnamefont{S.}~\bibnamefont{Stieberger}},
  \bibinfo{author}{\bibfnamefont{T.~R.} \bibnamefont{Taylor}},
  \bibnamefont{and} \bibinfo{author}{\bibfnamefont{B.}~\bibnamefont{Zhu}},
  \bibinfo{journal}{JHEP} \textbf{\bibinfo{volume}{09}}, \bibinfo{pages}{198}
  (\bibinfo{year}{2020}), \eprint{2007.03785}.

\bibitem[{\citenamefont{Kosower et~al.}(2019)\citenamefont{Kosower, Maybee, and
  O'Connell}}]{Kosower:2018adc}
\bibinfo{author}{\bibfnamefont{D.~A.} \bibnamefont{Kosower}},
  \bibinfo{author}{\bibfnamefont{B.}~\bibnamefont{Maybee}}, \bibnamefont{and}
  \bibinfo{author}{\bibfnamefont{D.}~\bibnamefont{O'Connell}},
  \bibinfo{journal}{JHEP} \textbf{\bibinfo{volume}{02}}, \bibinfo{pages}{137}
  (\bibinfo{year}{2019}), \eprint{1811.10950}.

\bibitem[{\citenamefont{Cristofoli et~al.}(2022)\citenamefont{Cristofoli,
  Gonzo, Kosower, and O'Connell}}]{Cristofoli:2021vyo}
\bibinfo{author}{\bibfnamefont{A.}~\bibnamefont{Cristofoli}},
  \bibinfo{author}{\bibfnamefont{R.}~\bibnamefont{Gonzo}},
  \bibinfo{author}{\bibfnamefont{D.~A.} \bibnamefont{Kosower}},
  \bibnamefont{and}
  \bibinfo{author}{\bibfnamefont{D.}~\bibnamefont{O'Connell}},
  \bibinfo{journal}{Phys. Rev. D} \textbf{\bibinfo{volume}{106}},
  \bibinfo{pages}{056007} (\bibinfo{year}{2022}), \eprint{2107.10193}.

\bibitem[{\citenamefont{Di~Vecchia et~al.}(2023)\citenamefont{Di~Vecchia,
  Heissenberg, Russo, and Veneziano}}]{DiVecchia:2023frv}
\bibinfo{author}{\bibfnamefont{P.}~\bibnamefont{Di~Vecchia}},
  \bibinfo{author}{\bibfnamefont{C.}~\bibnamefont{Heissenberg}},
  \bibinfo{author}{\bibfnamefont{R.}~\bibnamefont{Russo}}, \bibnamefont{and}
  \bibinfo{author}{\bibfnamefont{G.}~\bibnamefont{Veneziano}}
  (\bibinfo{year}{2023}), \eprint{2306.16488}.

\bibitem[{\citenamefont{Kosower et~al.}(2022)\citenamefont{Kosower, Monteiro,
  and O'Connell}}]{Kosower:2022yvp}
\bibinfo{author}{\bibfnamefont{D.~A.} \bibnamefont{Kosower}},
  \bibinfo{author}{\bibfnamefont{R.}~\bibnamefont{Monteiro}}, \bibnamefont{and}
  \bibinfo{author}{\bibfnamefont{D.}~\bibnamefont{O'Connell}},
  \bibinfo{journal}{J. Phys. A} \textbf{\bibinfo{volume}{55}},
  \bibinfo{pages}{443015} (\bibinfo{year}{2022}), \eprint{2203.13025}.

\bibitem[{\citenamefont{Buonanno et~al.}(2022)\citenamefont{Buonanno, Khalil,
  O'Connell, Roiban, Solon, and Zeng}}]{Buonanno:2022pgc}
\bibinfo{author}{\bibfnamefont{A.}~\bibnamefont{Buonanno}},
  \bibinfo{author}{\bibfnamefont{M.}~\bibnamefont{Khalil}},
  \bibinfo{author}{\bibfnamefont{D.}~\bibnamefont{O'Connell}},
  \bibinfo{author}{\bibfnamefont{R.}~\bibnamefont{Roiban}},
  \bibinfo{author}{\bibfnamefont{M.~P.} \bibnamefont{Solon}}, \bibnamefont{and}
  \bibinfo{author}{\bibfnamefont{M.}~\bibnamefont{Zeng}}, in
  \emph{\bibinfo{booktitle}{{Snowmass 2021}}} (\bibinfo{year}{2022}),
  \eprint{2204.05194}.

\bibitem[{\citenamefont{Travaglini et~al.}(2022)}]{Travaglini:2022uwo}
\bibinfo{author}{\bibfnamefont{G.}~\bibnamefont{Travaglini}}
  \bibnamefont{et~al.}, \bibinfo{journal}{J. Phys. A}
  \textbf{\bibinfo{volume}{55}}, \bibinfo{pages}{443001}
  (\bibinfo{year}{2022}), \eprint{2203.13011}.

\bibitem[{\citenamefont{Manohar et~al.}(2022)\citenamefont{Manohar, Ridgway,
  and Shen}}]{Manohar:2022dea}
\bibinfo{author}{\bibfnamefont{A.~V.} \bibnamefont{Manohar}},
  \bibinfo{author}{\bibfnamefont{A.~K.} \bibnamefont{Ridgway}},
  \bibnamefont{and} \bibinfo{author}{\bibfnamefont{C.-H.} \bibnamefont{Shen}},
  \bibinfo{journal}{Phys. Rev. Lett.} \textbf{\bibinfo{volume}{129}},
  \bibinfo{pages}{121601} (\bibinfo{year}{2022}), \eprint{2203.04283}.

\bibitem[{\citenamefont{Di~Vecchia et~al.}(2022)\citenamefont{Di~Vecchia,
  Heissenberg, and Russo}}]{DiVecchia:2022owy}
\bibinfo{author}{\bibfnamefont{P.}~\bibnamefont{Di~Vecchia}},
  \bibinfo{author}{\bibfnamefont{C.}~\bibnamefont{Heissenberg}},
  \bibnamefont{and} \bibinfo{author}{\bibfnamefont{R.}~\bibnamefont{Russo}},
  \bibinfo{journal}{JHEP} \textbf{\bibinfo{volume}{08}}, \bibinfo{pages}{172}
  (\bibinfo{year}{2022}), \eprint{2203.11915}.

\bibitem[{\citenamefont{Alessio and Di~Vecchia}(2022)}]{Alessio:2022kwv}
\bibinfo{author}{\bibfnamefont{F.}~\bibnamefont{Alessio}} \bibnamefont{and}
  \bibinfo{author}{\bibfnamefont{P.}~\bibnamefont{Di~Vecchia}},
  \bibinfo{journal}{Phys. Lett. B} \textbf{\bibinfo{volume}{832}},
  \bibinfo{pages}{137258} (\bibinfo{year}{2022}), \eprint{2203.13272}.

\bibitem[{\citenamefont{Zwanziger}(1973)}]{Zwanziger:1973if}
\bibinfo{author}{\bibfnamefont{D.}~\bibnamefont{Zwanziger}},
  \bibinfo{journal}{Phys. Rev. D} \textbf{\bibinfo{volume}{7}},
  \bibinfo{pages}{1082} (\bibinfo{year}{1973}).

\bibitem[{\citenamefont{Cachazo and Strominger}(2014)}]{Cachazo:2014fwa}
\bibinfo{author}{\bibfnamefont{F.}~\bibnamefont{Cachazo}} \bibnamefont{and}
  \bibinfo{author}{\bibfnamefont{A.}~\bibnamefont{Strominger}}
  (\bibinfo{year}{2014}), \eprint{1404.4091}.

\bibitem[{\citenamefont{Bern et~al.}(2014{\natexlab{a}})\citenamefont{Bern,
  Davies, Di~Vecchia, and Nohle}}]{Bern:2014vva}
\bibinfo{author}{\bibfnamefont{Z.}~\bibnamefont{Bern}},
  \bibinfo{author}{\bibfnamefont{S.}~\bibnamefont{Davies}},
  \bibinfo{author}{\bibfnamefont{P.}~\bibnamefont{Di~Vecchia}},
  \bibnamefont{and} \bibinfo{author}{\bibfnamefont{J.}~\bibnamefont{Nohle}},
  \bibinfo{journal}{Phys. Rev. D} \textbf{\bibinfo{volume}{90}},
  \bibinfo{pages}{084035} (\bibinfo{year}{2014}{\natexlab{a}}),
  \eprint{1406.6987}.

\bibitem[{\citenamefont{Bern et~al.}(2014{\natexlab{b}})\citenamefont{Bern,
  Davies, and Nohle}}]{Bern:2014oka}
\bibinfo{author}{\bibfnamefont{Z.}~\bibnamefont{Bern}},
  \bibinfo{author}{\bibfnamefont{S.}~\bibnamefont{Davies}}, \bibnamefont{and}
  \bibinfo{author}{\bibfnamefont{J.}~\bibnamefont{Nohle}},
  \bibinfo{journal}{Phys. Rev. D} \textbf{\bibinfo{volume}{90}},
  \bibinfo{pages}{085015} (\bibinfo{year}{2014}{\natexlab{b}}),
  \eprint{1405.1015}.

\bibitem[{\citenamefont{Comp\`ere et~al.}(2018)\citenamefont{Comp\`ere,
  Fiorucci, and Ruzziconi}}]{Compere:2018ylh}
\bibinfo{author}{\bibfnamefont{G.}~\bibnamefont{Comp\`ere}},
  \bibinfo{author}{\bibfnamefont{A.}~\bibnamefont{Fiorucci}}, \bibnamefont{and}
  \bibinfo{author}{\bibfnamefont{R.}~\bibnamefont{Ruzziconi}},
  \bibinfo{journal}{JHEP} \textbf{\bibinfo{volume}{11}}, \bibinfo{pages}{200}
  (\bibinfo{year}{2018}), \bibinfo{note}{[Erratum: JHEP 04, 172 (2020)]},
  \eprint{1810.00377}.

\bibitem[{\citenamefont{Donnay et~al.}(2020)\citenamefont{Donnay, Pasterski,
  and Puhm}}]{Donnay:2020guq}
\bibinfo{author}{\bibfnamefont{L.}~\bibnamefont{Donnay}},
  \bibinfo{author}{\bibfnamefont{S.}~\bibnamefont{Pasterski}},
  \bibnamefont{and} \bibinfo{author}{\bibfnamefont{A.}~\bibnamefont{Puhm}},
  \bibinfo{journal}{JHEP} \textbf{\bibinfo{volume}{09}}, \bibinfo{pages}{176}
  (\bibinfo{year}{2020}), \eprint{2005.08990}.

\bibitem[{\citenamefont{Donnay et~al.}(2022)\citenamefont{Donnay, Nguyen, and
  Ruzziconi}}]{Donnay:2022hkf}
\bibinfo{author}{\bibfnamefont{L.}~\bibnamefont{Donnay}},
  \bibinfo{author}{\bibfnamefont{K.}~\bibnamefont{Nguyen}}, \bibnamefont{and}
  \bibinfo{author}{\bibfnamefont{R.}~\bibnamefont{Ruzziconi}},
  \bibinfo{journal}{JHEP} \textbf{\bibinfo{volume}{09}}, \bibinfo{pages}{063}
  (\bibinfo{year}{2022}), \eprint{2205.11477}.

\bibitem[{\citenamefont{Agrawal et~al.}(2024)\citenamefont{Agrawal, Donnay,
  Nguyen, and Ruzziconi}}]{Agrawal:2023zea}
\bibinfo{author}{\bibfnamefont{S.}~\bibnamefont{Agrawal}},
  \bibinfo{author}{\bibfnamefont{L.}~\bibnamefont{Donnay}},
  \bibinfo{author}{\bibfnamefont{K.}~\bibnamefont{Nguyen}}, \bibnamefont{and}
  \bibinfo{author}{\bibfnamefont{R.}~\bibnamefont{Ruzziconi}},
  \bibinfo{journal}{JHEP} \textbf{\bibinfo{volume}{02}}, \bibinfo{pages}{120}
  (\bibinfo{year}{2024}), \eprint{2309.11220}.

\bibitem[{\citenamefont{Strominger}(2014)}]{Strominger:2013jfa}
\bibinfo{author}{\bibfnamefont{A.}~\bibnamefont{Strominger}},
  \bibinfo{journal}{JHEP} \textbf{\bibinfo{volume}{07}}, \bibinfo{pages}{152}
  (\bibinfo{year}{2014}), \eprint{1312.2229}.

\bibitem[{\citenamefont{He et~al.}(2015)\citenamefont{He, Lysov, Mitra, and
  Strominger}}]{He:2014laa}
\bibinfo{author}{\bibfnamefont{T.}~\bibnamefont{He}},
  \bibinfo{author}{\bibfnamefont{V.}~\bibnamefont{Lysov}},
  \bibinfo{author}{\bibfnamefont{P.}~\bibnamefont{Mitra}}, \bibnamefont{and}
  \bibinfo{author}{\bibfnamefont{A.}~\bibnamefont{Strominger}},
  \bibinfo{journal}{JHEP} \textbf{\bibinfo{volume}{05}}, \bibinfo{pages}{151}
  (\bibinfo{year}{2015}), \eprint{1401.7026}.

\bibitem[{\citenamefont{Strominger}(2017)}]{Strominger:2017zoo}
\bibinfo{author}{\bibfnamefont{A.}~\bibnamefont{Strominger}},
  \emph{\bibinfo{title}{{Lectures on the Infrared Structure of Gravity and
  Gauge Theory}}} (\bibinfo{year}{2017}), ISBN
  \bibinfo{isbn}{978-0-691-17973-5}, \eprint{1703.05448}.

\bibitem[{\citenamefont{Alessio and Arzano}(2019)}]{Alessio:2019cae}
\bibinfo{author}{\bibfnamefont{F.}~\bibnamefont{Alessio}} \bibnamefont{and}
  \bibinfo{author}{\bibfnamefont{M.}~\bibnamefont{Arzano}},
  \bibinfo{journal}{Phys. Rev. D} \textbf{\bibinfo{volume}{100}},
  \bibinfo{pages}{044028} (\bibinfo{year}{2019}), \eprint{1906.05036}.

\bibitem[{\citenamefont{Di~Vecchia et~al.}(2021)\citenamefont{Di~Vecchia,
  Heissenberg, Russo, and Veneziano}}]{DiVecchia:2021ndb}
\bibinfo{author}{\bibfnamefont{P.}~\bibnamefont{Di~Vecchia}},
  \bibinfo{author}{\bibfnamefont{C.}~\bibnamefont{Heissenberg}},
  \bibinfo{author}{\bibfnamefont{R.}~\bibnamefont{Russo}}, \bibnamefont{and}
  \bibinfo{author}{\bibfnamefont{G.}~\bibnamefont{Veneziano}},
  \bibinfo{journal}{Phys. Lett. B} \textbf{\bibinfo{volume}{818}},
  \bibinfo{pages}{136379} (\bibinfo{year}{2021}), \eprint{2101.05772}.

\bibitem[{\citenamefont{Alessio and Di~Vecchia}(2024)}]{Alessio:2024wmz}
\bibinfo{author}{\bibfnamefont{F.}~\bibnamefont{Alessio}} \bibnamefont{and}
  \bibinfo{author}{\bibfnamefont{P.}~\bibnamefont{Di~Vecchia}}
  (\bibinfo{year}{2024}), \eprint{2402.06533}.

\bibitem[{\citenamefont{Lippstreu}(2023)}]{Lippstreu:2023vvg}
\bibinfo{author}{\bibfnamefont{L.}~\bibnamefont{Lippstreu}}
  (\bibinfo{year}{2023}), \eprint{2312.08455}.

\bibitem[{\citenamefont{McLoughlin et~al.}(2022)\citenamefont{McLoughlin, Puhm,
  and Raclariu}}]{McLoughlin:2022ljp}
\bibinfo{author}{\bibfnamefont{T.}~\bibnamefont{McLoughlin}},
  \bibinfo{author}{\bibfnamefont{A.}~\bibnamefont{Puhm}}, \bibnamefont{and}
  \bibinfo{author}{\bibfnamefont{A.-M.} \bibnamefont{Raclariu}},
  \bibinfo{journal}{J. Phys. A} \textbf{\bibinfo{volume}{55}},
  \bibinfo{pages}{443012} (\bibinfo{year}{2022}), \eprint{2203.13022}.

\bibitem[{\citenamefont{Pasterski and Puhm}(2021)}]{Pasterski:2020pdk}
\bibinfo{author}{\bibfnamefont{S.}~\bibnamefont{Pasterski}} \bibnamefont{and}
  \bibinfo{author}{\bibfnamefont{A.}~\bibnamefont{Puhm}},
  \bibinfo{journal}{Phys. Rev. D} \textbf{\bibinfo{volume}{104}},
  \bibinfo{pages}{086020} (\bibinfo{year}{2021}), \eprint{2012.15694}.

\bibitem[{\citenamefont{Lam and Shao}(2018)}]{Lam:2017ofc}
\bibinfo{author}{\bibfnamefont{H.~T.} \bibnamefont{Lam}} \bibnamefont{and}
  \bibinfo{author}{\bibfnamefont{S.-H.} \bibnamefont{Shao}},
  \bibinfo{journal}{Phys. Rev. D} \textbf{\bibinfo{volume}{98}},
  \bibinfo{pages}{025020} (\bibinfo{year}{2018}), \eprint{1711.06138}.

\bibitem[{\citenamefont{Witten}(1998)}]{Witten:1998qj}
\bibinfo{author}{\bibfnamefont{E.}~\bibnamefont{Witten}},
  \bibinfo{journal}{Adv. Theor. Math. Phys.} \textbf{\bibinfo{volume}{2}},
  \bibinfo{pages}{253} (\bibinfo{year}{1998}), \eprint{hep-th/9802150}.

\bibitem[{\citenamefont{Costa et~al.}(2014)\citenamefont{Costa, Gon\c{c}alves,
  and Penedones}}]{Costa:2014kfa}
\bibinfo{author}{\bibfnamefont{M.~S.} \bibnamefont{Costa}},
  \bibinfo{author}{\bibfnamefont{V.}~\bibnamefont{Gon\c{c}alves}},
  \bibnamefont{and} \bibinfo{author}{\bibfnamefont{J.~a.}
  \bibnamefont{Penedones}}, \bibinfo{journal}{JHEP}
  \textbf{\bibinfo{volume}{09}}, \bibinfo{pages}{064} (\bibinfo{year}{2014}),
  \eprint{1404.5625}.

\bibitem[{\citenamefont{Banerjee}(2019)}]{Banerjee:2018gce}
\bibinfo{author}{\bibfnamefont{S.}~\bibnamefont{Banerjee}},
  \bibinfo{journal}{JHEP} \textbf{\bibinfo{volume}{01}}, \bibinfo{pages}{205}
  (\bibinfo{year}{2019}), \eprint{1801.10171}.

\bibitem[{\citenamefont{Lippstreu}(2021)}]{Lippstreu:2021avq}
\bibinfo{author}{\bibfnamefont{L.}~\bibnamefont{Lippstreu}},
  \bibinfo{journal}{JHEP} \textbf{\bibinfo{volume}{11}}, \bibinfo{pages}{051}
  (\bibinfo{year}{2021}), \eprint{2106.00084}.

\bibitem[{\citenamefont{Alessio and Esposito}(2018)}]{Alessio:2017lps}
\bibinfo{author}{\bibfnamefont{F.}~\bibnamefont{Alessio}} \bibnamefont{and}
  \bibinfo{author}{\bibfnamefont{G.}~\bibnamefont{Esposito}},
  \bibinfo{journal}{Int. J. Geom. Meth. Mod. Phys.}
  \textbf{\bibinfo{volume}{15}}, \bibinfo{pages}{1830002}
  (\bibinfo{year}{2018}), \eprint{1709.05134}.

\bibitem[{\citenamefont{Law and Zlotnikov}(2020{\natexlab{b}})}]{Law:2019glh}
\bibinfo{author}{\bibfnamefont{Y.~T.~A.} \bibnamefont{Law}} \bibnamefont{and}
  \bibinfo{author}{\bibfnamefont{M.}~\bibnamefont{Zlotnikov}},
  \bibinfo{journal}{JHEP} \textbf{\bibinfo{volume}{03}}, \bibinfo{pages}{085}
  (\bibinfo{year}{2020}{\natexlab{b}}), \bibinfo{note}{[Erratum: JHEP 04, 202
  (2020)]}, \eprint{1910.04356}.

\bibitem[{\citenamefont{Stieberger and Taylor}(2019)}]{Stieberger:2018onx}
\bibinfo{author}{\bibfnamefont{S.}~\bibnamefont{Stieberger}} \bibnamefont{and}
  \bibinfo{author}{\bibfnamefont{T.~R.} \bibnamefont{Taylor}},
  \bibinfo{journal}{Phys. Lett. B} \textbf{\bibinfo{volume}{793}},
  \bibinfo{pages}{141} (\bibinfo{year}{2019}), \eprint{1812.01080}.

\bibitem[{\citenamefont{Oblak}(2015)}]{Oblak:2015qia}
\bibinfo{author}{\bibfnamefont{B.}~\bibnamefont{Oblak}} (\bibinfo{year}{2015}),
  \eprint{1508.00920}.

\end{thebibliography}

\end{document}